\documentclass[11pt,a4paper]{article}
\pdfoutput=1

\usepackage[]{Packages}

\usepackage{Definitions}

\usepackage{pifont}

\bibliography{unifiedcites}

\allowdisplaybreaks

\newcommand{\OfficialTitle}{Extended Gauge Theory Deformations From~Flux~Backgrounds}

\title{%
  \Huge
  \textbf{
      \dosserif
      \OfficialTitle
    }  
}

\hypersetup{pdfauthor={Neil Lambert and Domenico Orlando and Susanne Reffert and Yuta Sekiguchi},pdftitle={\OfficialTitle}}

\author{%
  \begin{minipage}{.9\linewidth}
    \vspace{1cm}
    \begin{center}
      {\small \textbf{Neil Lambert}\textsuperscript{\ding{72}}, \textbf{Domenico Orlando}\textsuperscript{\ding{74}\ding{73}}, \textbf{Susanne Reffert}\textsuperscript{\ding{73}}, and \textbf{Yuta Sekiguchi}\textsuperscript{\ding{73}}}
    \end{center}
    \vspace{1cm}
    \authorBlock{\ding{72}}{Department of Mathematics \\ King's College London \\ The Strand, London, WC2R 2LS, UK}
    \authorBlock{\ding{74}}{INFN sezione di Torino and Arnold--Regge Center\\ via Pietro Giuria 1, 10125 Turin, Italy}
    \authorBlock{\ding{73}}{Albert Einstein Center for Fundamental Physics\\ Institute for Theoretical Physics,  University of Bern,\\  Sidlerstrasse 5, \textsc{ch}-3012 Bern, Switzerland}
  \end{minipage}
}

\date{}

\begin{document}

\setstretch{1.2}

\numberwithin{equation}{section}

\begin{titlepage}

  \maketitle

  \thispagestyle{empty}

  \vfill
  \abstract
  
We consider  supersymmetric deformations of gauge theories in various dimensions obtained from a String Theory realisation of branes embedded in flux backgrounds. In particular we obtain deformations which take the form of Wilson line defects, where the R-symmetry is twisted into the gauge symmetry. Furthermore we construct higher-order generalisations, also expressed a twisting of the R-symmetry, that have symmetries associated to co-dimension two and three defects. 
  \vfill

\end{titlepage}

\tableofcontents

\section{Introduction}
\label{sec:introduction}

Ever since the pioneering work of Nekrasov~\cite{Nekrasov:2002qd} there has been great interest in studying supersymmetric but non-Lorentz invariant deformations of gauge theories.  In this paper our
 approach to studying deformations and defects in supersymmetric gauge theories consists of realising the whole setup in string theory.
The string-theoretical picture consists of branes embedded in a flux background and allows us to access and relate a variety of systems with deformations and defects via duality cascades including T-dualities, S-dualities and lifts.
In particular we realise the gauge theory in terms of the fluctuations of a D~brane using a \ac{dbi} action.
The presence of the deformation is encoded by fluxes which are switched on in the bulk spacetime where the brane is embedded. When the fluxes are misaligned, {\it i.e.} break part of the symmetry preserved by the brane setup, the resulting gauge theory receives a deformation. 
In this way the  Omega-deformation of~\cite{Nekrasov:2002qd}  can be constructed as a brane in the fluxtrap background~\cite{Hellerman:2011mv, Reffert:2011dp, Orlando:2013yea}. However 
whole families of deformations can be realised in this way and we wish to investigate some of these here. 

We note that
recently a similar approach to  constructing deformations of 4-dimensional ${\cal N}=4$ super-Yang-Mills has appeared in~\cite{Choi:2018fqw, Choi:2017kxf}. In those works the focus was on constructing theories with variable couplings by placing \D3-branes in backgrounds generated by other \D{p}-branes. Here we describe deformations which we associate to a twisting of the gauge theory connection with the R-symmetry and which are valid in a variety of dimensions. In particular we consider a different class of flux backgrounds which involve a choice of anti-symmetric 2-tensor $\omega^{IJ}$ transverse to the branes that we identify with a generator of the R-symmetry.

This brane approach is in many ways technically simpler than direct gauge theory calculations. The string construction gives a ten-dimensional geometric perspective that makes the symmetries of the system more manifest and easily accessible. In particular, the supersymmetric properties are very easily described.
Using the \ac{dbi} action, and the String Theory realisation,  we have a fully microscopic description that goes beyond the information contained in the partition function.
We also give the fermionic actions and the supersymmetry transformations derived from the branes, both for the Abelian and the non-Abelian cases.

\medskip
The curved string theory background that we use in this work is S-dual to the fluxtrap deformation discussed in~\cite{Hellerman:2011mv, Reffert:2011dp, Hellerman:2012zf, Lambert:2013lxa, Lambert:2014fma}.
This results in the presence of \ac{rr} fields instead of a \(B\) field.
Another difference to earlier constructions is that all the spatially extended branes discussed here are extended in the  "Melvin" direction, which plays a special role for the background deformation. 
Unlike in previous examples of gauge theory deformations stemming from a flux background in string theory, where the deformations took the form of (twisted) mass deformations or Omega-type deformations, we find here deformations of a \emph{different form}. %
In the simplest case of the construction discussed here  the deformation has an obvious interpretation as a  Wilson line defect where the R-symmetry is twisted into the gauge symmetry.  However we will also present higher-dimensional analogues whose interpretation as a defect is less clear-cut but which also involve a gauge connection that is twisted by the R-symmetry.

\bigskip
The plan of this article is as follows. In Section~\refstring{sec:line-defects}, as a warm-up, we discuss first a particle propagating in a magnetic field in terms of a Wilson line for a gauge connection that is twisted by the R-symmetry.
We then realise this in String Theory as a \D0~brane in a \ac{rr} flux background. Via T--duality, we can reach general \D{p}~branes, giving rise to higher-dimensional gauge theories with line defects (Section.~\ref{sec:lift-Dp}).
We finally generalise our treatment to non-Abelian gauge theories and discuss the conserved supersymmetries in Section~\ref{sec:susy-nonAb}. In Section~\ref{sec:wilson-lines}, we give a general discussion of Wilson lines of global symmetries.
In Section~\refstring{sec:lift-M}, we discuss the lift to M-theory and study   deformed \ac{blg} and \ac{abjm} models, including a maximally supersymmetric deformation of \ac{abjm} theory.

Via a duality cascade, we can reach various \D{}~brane configurations giving rise to novel deformations which are reminiscent of higher-dimensional defects. In Section~\refstring{sec:higher-dimensional-defects}, we discuss higher-dimensional deformations starting from a background which is related to the one used for the \D1~brane case  by two T--dualities. We start with the case of a \D5~brane in this background, as it is the simplest, only containing a $C_4$ form and no dilaton (Sec.~\ref{sec:D5}). It gives rise to a deformation of a 6-dimensional gauge theory that has a natural interpretation as two orthogonal  3-dimensional defects. We give the supersymmetric \D5~brane action, discuss the global symmetries of this configuration, and the equations of motion for the scalars and gauge field. We find that the supersymmetry transformations are modified by the deformation (to first order in the deformation parameter) such that the spinorial transformation parameter receives  space-dependent correction. Finally, we discuss the non-Abelian generalisation of this case which arises for a stack of multiple \D5~branes.

Another interesting case, related to the one of the \D5 by two T-dualities, is the one arising from a \D3~brane which we study in section \ref{sec:D3}. This results in a deformed four-dimensional gauge theory with a natural interpretation as two orthogonal 2-dimensional defects. The discussion follows the same lines as the one of the \D5~brane case treated before. 
All the other cases which can be reached from the \D5~brane case via T-dualities are summarised in Section~\ref{sec:Duality-Cascade}.

In Section~\refstring{sec:conclusions} we present concluding remarks.
In Appendix~\ref{sec:Melvin}, the fluxtrap construction is reviewed and in Appendix~\ref{sec:notation} we give our conventions for the notation.

\section{Deformations and line defects}
\label{sec:line-defects}

In this section, we consider a \emph{twisted Wilson line defect}. A time-like Wilson line, as in our case, is also known as a Polyakov line in the literature.

\subsection{Particle in a magnetic field}
\label{sec:D0}

Let us start with the simplest physical system of the class we are aiming to describe here: a massless complex scalar field \(Z(t)\) in zero dimensions.
The system has a manifest \(U(1)\) symmetry that rotates \(Z(t) \mapsto e^{i \lambda} Z(t)\) that we gauge with a field \(A(t)\).
The Lagrangian is given by
\begin{equation}
  L = \frac{1}{2g^2} \pqty{ \del_0 + A(t)} Z(t) \pqty{ \del_0 + \bar A(t)} \bar Z(t) ,
\end{equation}
and it is clearly invariant under
\begin{equation}
  \begin{cases}
    Z(t) \mapsto e^{i\lambda(t)} Z(t) \\
    A(t) \mapsto A(t) - i \del_0 \lambda(t) .
  \end{cases}
\end{equation}
If we give to \(A\) a large constant \ac{vev}, \(\ev{A} = i \varepsilon\), the action reads
 (\(Z = X + i Y\)) 
 \begin{equation}
  \label{eq:scalar-field-action}
  \begin{aligned}
    L &= \frac{1}{2g^2} \pqty{ \del_0 + i \varepsilon} Z(t) \pqty{ \del_0 - i \varepsilon} \bar Z(t)  \\
    &= \frac{1}{2g^2} \bqty{ \dot X(t)^2 + \dot Y(t)^2 + 2 \varepsilon X(t) \dot Y(t) - 2 \varepsilon Y(t) \dot X(t) + \varepsilon^2 \pqty{X(t)^2 + Y(t)^2} }.
  \end{aligned}
\end{equation}
This is the same as the action of a classical particle of unit mass and charge in two dimensions propagating in the presence of a constant magnetic field orthogonal to the plane (with potential \(A = \varepsilon y \dd{x} - \varepsilon x \dd{y}\)) and with a scalar potential \(V(x,y) = - \frac{\varepsilon^2}{2g^{2}} \pqty{ x^2 + y^2} \).

The corresponding \ac{eom} is
\begin{equation}
  \ddot Z(t) - 2 i \varepsilon \dot Z(t) - \varepsilon^2 Z = 0 ,\\
\end{equation}
which admits the general solution
\begin{equation}
  Z (t) = \pqty{\rho + C t} \eu^{-i \varepsilon t},
\end{equation}
where \(\rho \) and \(C\) are real constants.

Performing a Legendre transform we see that \(\varepsilon\) takes the role of a chemical potential for the \(U(1)\) rotation acting linearly on \(Z\),
\begin{equation}
  H =  2 g^2  P  \bar P - i \varepsilon \pqty{ Z P - \bar Z \bar P }.
\end{equation}
The energy of the classical solution is \(H = \frac{1}{2g^2} C^2\) and it is minimal for \(C = 0\).
This gives us an intuitive picture of the dynamics.
The particle moves in a circle of radius $\rho$ in the complex plane with constant angular velocity \(\epsilon\).

This very simple example admits many generalisations.
For instance, we can have multiple charged scalars \(Z^A\) and add fermions to make the system supersymmetric.
Another possibility is to go to higher dimensions and write a field theory in \(d+1\) dimensions.
In this spirit, we want to embed our construction in string theory.
Then the field \(Z^A\) encodes the fluctuations of a \D0~brane moving in the direction \(Z^A\).
The \ac{vev} of the gauge field is realised in terms of a flux in the bulk and the potential term arises from gravitational back reaction of the flux. 
In this language, neglecting the fluctuations of the field \(A\) amounts to taking the probe limit for the dynamics of the \D0~brane.
The \ac{vev} is nothing else than the pullback of the (non-dynamical) bulk \ac{rr} field in the \ac{dbi} action.

\subsection{Twisted Wilson line for \D0~branes}
\label{sec:D1}

Next, we consider a string theory set-up where a \D0~brane is extended along $x^0$, while an \ac{rr} flux background deforms the directions $x^1,\dots,x^8$  orthogonal to the brane.
At first order in the deformation, we have a constant two-form flux orthogonal to the brane.
The flux defines an element of the ${\mathfrak so}(8)$ R--symmetry algebra.
Due to the standard \ac{wz} coupling to the background \ac{rr}--flux one finds that  the flux induces a twist of the  gauge theory $\mathfrak{u}(N)$ connection with the R--symmetry.
The flux manifests itself in the gauge theory as a background Wilson line in this twisted connection.

\paragraph{The string theory background.}

Take the \tIIB background that we refer to as the \ac{rr} fluxtrap, which was introduced in~\cite{Lambert:2014fma} as an S-dual of the fluxtrap solution~\cite{Hellerman:2011mv} (see Appendix~\ref{sec:Melvin} for details). The bulk fields are given by
\begin{equation}
  \label{eq:fluxtrap}
  \begin{aligned}
    \dd s^2_{10} &=  \Delta \left[ - (\dd x^0)^2+ (\dd x^1)^2 + \left(\delta_{IJ}- \frac {U_I U_J}{\Delta^2}\right)  \dd x^I  \dd x^J   \right] \, ,\\
    \Phi & = \frac{3}{2} \log\Delta \,,\\
    C_{1} &= \frac{1}{\Delta^2} U\,,
  \end{aligned}
\end{equation}
where
\begin{equation}\label{eq:U}
\begin{aligned}
  U & = U_J\dd x^J\\
  & = \frac{1}{2}\omega_{IJ}x^I dx^J\\
  &= \frac{i}{4}\sum_{A=1}^{4} \varepsilon_{A} (z^{A}\dd \bar z^{A} - \bar z^{A}\dd z^{A}),
\end{aligned}\end{equation}
and
\begin{equation}\label{eq:Delta}
  \Delta = \sqrt{1 + U_I U_J \delta^{IJ}}\,.
\end{equation}
Here, and in the following, we have introduced both real coordinates $x^I$, $I=2,...,9$ and complex coordinates 
\begin{equation}
z^{A} =  {x^{2A}+ix^{2A+1}} \ ,\qquad A=1,2,3,4 
\end{equation}
which  diagonalise  $i\omega_{IJ}$ with eigenvalues $\pm  \varepsilon_A$. 
Note that in the original construction of the fluxtrap as a T-dual of Melvin space (also known as a fluxbrane background) the coordinate $x^1$ was taken to be periodic (the Melvin direction, see Appendix~\ref{sec:Melvin}). However once this solution is obtained one can allow the periodicity to be arbitrary or even infinite as we do here.

To first order in the deformation parameters $\varepsilon_A$ this solution is simply a  constant two-form \ac{rr} flux in string theory in a flat background. The full solution includes the complete gravitational back-reaction to all orders in the parameters $\varepsilon_A$. The discussion  here is therefore similar in spirit to that of~\cite{Lambert:2009qw}. In that paper M2~branes were placed in a flux background which, to first order in the fluxes, is flat and preserves supersymmetry but induces mass-like deformations on the M2~brane gauge theory. However supersymmetry requires that there must also be second-order corrections to the gauge theory and these can be interpreted as arising from spacetime curvature due to the gravitational back reaction of the fluxes.
A key difference here is that  (\ref{eq:fluxtrap}) is the full back reacted solution and therefore the \ac{dbi} action captures all the necessary supersymmetric deformations to the brane, at least in the Abelian case. 
Thus in this paper we will use the \ac{dbi} action to construct the deformed Abelian brane theory. We will then explicitly construct the supersymmetry and find the non-Abelian extension.

\bigskip

The number of preserved supersymmetries is determined by the equation~\cite{Hellerman:2012zf}
\begin{equation}
  \omega_{IJ} \Gamma^{IJ} \epsilon = 2i\sum_{A=1}^4 \varepsilon_{A} \Gamma^{2A(2A+1)} \epsilon = 0  ,
\end{equation}
where \(\epsilon \) is a ten-dimensional chiral spinor.
The following alternatives are possible:
\begin{itemize}
\item for general values of deformation parameters $\{\varepsilon_A\}_{A=1}^{4}$, all supersymmetries are broken;
\item for \(\sum_{A=1}^{4}\varepsilon_A=0\), some of the Killing spinors are preserved. Each independent non-vanishing \(\varepsilon\) reduces the supersymmetry by one half;
\item for \( \varepsilon_A = \varepsilon \), \(\forall A = 1,2,3,4\), remarkably, there are \emph{twelve linearly independent Killing spinors}. %
\item for $\varepsilon_{1}=\varepsilon_{2}$ and $\varepsilon_{3} = \varepsilon_{4}$, eight supercharges are unbroken. %
\end{itemize}
\paragraph{Gauge theory action in two dimensions.}

Consider now a \D0~brane in this background, extended in the direction $x^0 $ as given in Table~\ref{tab:D1}.
\begin{table}
  \centering
  \begin{tabular}{lcccccccccc}
    \toprule
    \( x \)   & 0               & 1               & 2               & 3  & 4  & 5  & 6  & 7 & 8  & 9  \\
    \midrule
    fluxtrap& &   &\ep{\varepsilon_1} & \ep{\varepsilon_2} & \ep{\varepsilon_3} &  \ep{\varepsilon_4}                  \\
    \D0~brane & \X & &  \ep{Z^1} & \ep{Z^2} & \ep{Z^3} &   \ep{Z^4}   \\
        \bottomrule
  \end{tabular}
 \caption{\D0~brane and its scalar fields in the fluxtrap background}  \label{tab:D1}
\end{table}
The bosonic action for the static embedding of the brane is
\begin{multline}
  S_{\D0} = -\frac{1}{2g^2} \int \dd{x^0} \Bigg[ \sum_{I=1}^9   \del_0 X^I \del^0  X^I  + \sum_{I,J,K}\omega_{IK}\omega_{JK} X^JX^K    - 2\sum_{IJ}\omega_{IJ} X^I\del_0 X^J  \Bigg].
\end{multline}
In order to facilitate the interpretation, we introduce complex coordinates and rewrite the action as
\begin{equation}
  \label{eq:D0-action}
  S_{\D0} = -\frac{1}{2 g^2} \int \dd{x^0}\bqty{\del_0 X^1 \del^0 X^1 + \sum_{A = 1}^4 \pqty{\del_0 Z^A+ i\varepsilon_A  Z^A } \pqty{\del^0 \bar Z^A -i   \varepsilon_A  \bar Z^A}} ,
\end{equation}
where the \(Z^A\) are the complex fields in Table~\ref{tab:D1}.
The contribution at first order in \(\varepsilon\) comes from the \ac{rr} flux in the bulk via the \ac{cs} term; the metric and the dilaton contribute to the quadratic term.
As expected, this is the generalisation of the action in Eq.~\eqref{eq:scalar-field-action} to more than one field.
In this case, moreover, the system is supersymmetric and the fermionic part of the action is  readily evaluated using the results of~\cite{Marolf:2003vf} and it is given by
\begin{equation}
  \label{eq:D0-fermion}
  \begin{aligned}
    S_F &=  \frac{\im}{2 g^2} \int \dd{x^0} \bar \Psi \hat \Gamma^0 \del_0 \Psi + \sum_{A=1}^4 \bar \Psi \varepsilon_A  \Gamma^0 \Gamma^{2A (2A+1)} \Psi \\
    & = \frac{\im}{2 g^2} \int \dd{x^0} \bar \Psi \hat \Gamma^0 \pqty{\del_0 + i  \sum_{A=1}^4 \varepsilon_A    \Gamma^{2A (2A+1)}} \Psi\,,
  \end{aligned}
\end{equation}
where \(\Psi\) is a \(32\)--dimensional Majorana spinor subject to the constraint $   \Gamma_{11}\Psi=-\Psi $.
The \D0~brane will break half of the supercharges preserved by the bulk fields.
We will return to this in the following, when we discuss the higher-dimensional generalisations of this construction. We have therefore succeeded in realising the simple case of a particle in a magnetic field, which was discussed at the beginning of this section,  as a \D0~brane in String Theory.

\subsection{The Dp~branes}
\label{sec:lift-Dp}

Setting various $\varepsilon$'s of our original background to zero results in isometries along the associated directions. We can therefore T-dualise in those directions to obtain \D{p}~branes in a background flux.

\paragraph{Action in higher dimensions.}

The $(p+1)$-form potential then takes the form (up to second order in $\varepsilon$):
\begin{equation}
  \label{eq:C-p+1}
  \begin{aligned}
    C^{(p+1)} &= \sum_{IJ} \frac 12 \omega_{IJ}x^I\dd x^J   \wedge \dd x^{10-p} \wedge \dots \wedge \dd x^9 \\
    & = U\wedge  \dd x^{10-p} \wedge \dots \wedge \dd x^9 .
  \end{aligned}
\end{equation}
The corresponding flux is $G_{p+2} = \dd C_{p+1}$.
Let us now introduce a \D{p}~brane that is mis-aligned, in the sense that $G_{p+2}$ has two indices off the brane.
We use the coordinates $x^\mu$, $\mu = 0,\dots,p$ along the brane and $I,J=p+1,p+2,\dots$.
From the \ac{cs} term there is a coupling (a hat denotes the pull-back to the world volume),
\begin{equation}\label{eq:cs_Abelian}
  \begin{aligned}
    S_{cs} &=   \int \hat C_{p+1} \\
    &= \int \dd[p+1]{x} \varepsilon^{\mu_0..\mu_p}G_{\mu_0...\mu_{p-1} IJ}X^I\partial_{\mu_{p}} X^J \\
    & = \int \dd[p+1]{x} \xi^\lambda \omega_{IJ} X^I\partial_{\lambda} X^J,
  \end{aligned}
\end{equation}
where we have rewritten $G_{p+1}=\star \xi_1\wedge \omega$, with $\omega=\dd U$ a 2-form transverse to the brane and $\xi_1$ a one-form along the brane.

The 2-form $\omega$ can be thought of as an element of the R--symmetry Lie algebra and as such generates a rotation in the transverse space. Spinors which satisfy
\begin{equation}
  \label{omegasusy}
  \omega_{IJ}\Gamma^{IJ}\epsilon =0
\end{equation}
are preserved by the rotation and therefore one might expect that the brane preserves these supersymmetries under the deformation induced by the flux.

The form of the action is the natural generalisation of the one in Eq.~\eqref{eq:D0-action}:
\begin{equation}
  S_{\D{p}} = -\frac{1}{2 g^2} \int \dd[2]{x}\bqty{ \frac{1}{2} F^2 + \sum_{A = 1}^{\floor{(9-p)/2}} \pqty{\del_\mu Z^A+ i\varepsilon_A \xi_\mu Z^A } \pqty{\del^\mu \bar Z^A -i   \varepsilon_A \xi^\mu \bar Z^A}} ,
\end{equation}
where 
\begin{equation}
 \xi_\mu  = \delta\indices{_\mu^0}
\end{equation}
is a constant Polyakov line \emph{i.e.} represents a constant \ac{vev} for the \(U(1)\) gauge field as in \S~\ref{sec:D0}.
We find that the presence of the flux in the bulk is translated into a non-trivial background value of a Polyakov line in the gauge theory that describes the motion of the \D{p}~brane.
In other words, the undeformed theory is coupled to a one-dimensional defect.
The fermionic term in Eq.~\eqref{eq:D0-fermion} is also directly generalised to obtain a supersymmetric action, as discussed in the next section.

\subsection{Supersymmetry and non-Abelian generalisation}\label{sec:susy-nonAb}

The analysis of the previous section can be repeated for the non-Abelian configuration of a stack of \D{p}~branes.
For the \ac{dbi} part once more we see that the metric does not contribute to terms in \(\varepsilon\) if we limit ourselves to terms with two derivatives.
The only new contribution comes from the dilaton and it has the same form as for the Abelian case.
We have:
\begin{multline}
  S_{DBI} =  -\frac{1}{ g^2} \Tr \int \dd[p+1]{x} \Bigg[ \frac{1}{4} F^2 + \frac{1}{2} \del_\mu X^I \del^\mu  X^I  +   \frac{1}{2}\omega_{IK}\omega_{JK} X^IX^J\\ - \frac{1}{4} \comm{X^I}{X^J} \comm{X^I}{X^J}  \Bigg],
\end{multline}
where now \(X^I\) is a matrix and $D$ is a Lie-algebra valued covariant derivative.
Following~\cite{Myers:1999ps}, the \ac{cs} term for \D{}~branes in a generic \ac{rr}-field background is
\begin{equation}\label{eq:CS}
  S_{CS} = \frac{1}{g^2} \Tr \int \bqty{ \pqty{\eu^{2 \pi i \imath_X \imath_X} \sum_n C^{(n)}} \eu^{2 \pi i F} },
\end{equation}
where \(\imath_X\) is the interior product with \(X^I\) seen as a vector in the transverse space.
The only relevant term in our configuration corresponds to the field \(C^{[p+1]} = U \wedge \dd{x^{10-p}} \wedge \dots \wedge \dd{x^{9}}\) in Eq.~\eqref{eq:C-p+1}.
The field has \(p\) legs in the worldvolume of the brane and only one in the transverse space.
It follows that no contraction is possible with \({X}^I {X}^J\) and the only term remaining is the natural generalisation of Eq.~\eqref{eq:cs_Abelian}
\begin{equation}
  \begin{aligned}
  S_{CS} %
  &= \frac{1}{g^2}\int \dd[p+1]{x} \xi^\lambda \omega_{IJ}\Tr (X^ID_\lambda X^J).
\end{aligned}
\end{equation}

\paragraph{Supersymmetry.}
We start with the usual (undeformed) action for a \D{p}~brane, given by ten-dimensional \ac{sym} theory reduced to $p+1$ dimensions:
\begin{multline}
    S_{SYM} = -\frac{1}{g^2}\Tr\int \dd[p+1]{x} \Bigg[\frac{1}{4}F_{\mu\nu}F^{\mu\nu}+\frac{1}{2}D_\mu X^I D^\mu X^I  +\frac{i}{2}\bar\Psi\Gamma^\mu D_\mu \Psi \\
    + \frac{1}{2}\bar\Psi\Gamma^I \comm{X^I}{\Psi}-\frac{1}{4} \comm{X^I}{X^J} \comm{X^I}{X^J}  \Bigg]  .
\end{multline}
We use a notation where $\Psi$ and $\epsilon$ are 32-component Majorana spinors subject to the constraints
\begin{align}
  \Gamma_{11}\Psi &= -\Psi,& \Gamma_{11}\epsilon &= \epsilon .
\end{align}
This action is invariant under the supersymmetry transformations
\begin{equation}
  \begin{aligned}
    \delta X^I & = i\bar\epsilon \Gamma^I\Psi,\\
    \delta A_\mu & = i\bar\epsilon \Gamma_\mu\Psi, \\
    \delta \Psi & = \Gamma^\mu\Gamma^I{ D}_\mu X^I +
    \frac{1}{2}\Gamma^{\mu\nu}F_{\mu\nu}\epsilon
    -\frac{i}{2}\Gamma^{IJ}[X^I,X^J]\epsilon\ .
  \end{aligned}
\end{equation}
The deformation arises from the flux term
\begin{equation}
  S_{\text{cs}} =  \frac{1}{g^2} \Tr \int \dd[p+1]{x} \xi^\lambda \omega_{IJ} \bqty{X^ID_\lambda X^J-\frac{i}{8}\bar\Psi\Gamma_\lambda\Gamma^{IJ}\Psi} ,
\end{equation}
where we have made the fields non--Abelian and also included a fermionic term. Here we see that $S_{\text{cs}}$ can be written as
\begin{equation}
S_{\text{cs}} =  \frac{1}{g^2}\int \xi^\lambda \omega_{IJ}J^{IJ}_\lambda,
\end{equation}
where $J^{IJ}_\lambda$ is the R--symmetry current. In addition to this linear perturbation in  the flux there is a gravitational back-reaction which induces a second-order term
\begin{align}
  S_{2} = -\frac{1}{2g^2}\int \xi^\lambda \xi_\lambda \omega_{IK}\omega_{JK} \Tr (X^IX^J) .
\end{align}
These terms can all be deduced in the Abelian case by examining the standard brane action including \ac{cs} terms.

One then sees that the resulting action
\begin{align}
S = S_{SYM}+S_{\text{cs}}+S_2
\end{align}
 simply corresponds to making the replacement
\begin{equation}\label{defD}
  \begin{aligned}
    D_\mu X^I &\to \mathcal{D}_\mu X^I = D_\mu X^I + \xi_\mu \omega_{IK}X^K, \\
    D_\mu \Psi &\to \mathcal{D}_\mu \Psi = D_\mu \Psi + \frac{1}{4}\xi_\mu
    \omega_{KL}\Gamma^{KL}\Psi
  \end{aligned}
\end{equation}
in the original \ac{sym} action.
Thus the effect of the flux is to induce a \emph{twisting of the world volume gauge symmetry with the R--symmetry}.
The deformed action is invariant under the supersymmetry transformation where $D_\mu$ is now replaced by $\mathcal{D}_\mu$:
\begin{equation}
  \begin{aligned}
    \delta X^I & = i\bar\epsilon \Gamma^I\Psi, \\
    \delta A_\mu & = i\bar\epsilon \Gamma_\mu\Psi, \\
    \delta \Psi & = \Gamma^\mu\Gamma^I\mathcal{D}_\mu X^I\epsilon +
    \frac{1}{2}\Gamma^{\mu\nu}F_{\mu\nu}\epsilon
    -\frac{i}{2}\Gamma^{IJ} \comm{X^I}{X^J}\epsilon ,
  \end{aligned}
\end{equation}
provided that
\begin{align}
  \label{econ-D-branes}
  \Gamma_{11} \epsilon &=\epsilon  ,& \omega_{IJ}\Gamma^{IJ}\epsilon &=0 .
\end{align}
Furthermore one sees that the algebra closes onto translations, gauge transformations and R-symmetries:
\begin{equation}
\begin{aligned}
[\delta_1,\delta_2]X^I  &= 2i(\bar\epsilon_2\Gamma^\nu\epsilon_1){\cal D}_\nu X^I    \ ,\\
[\delta_1,\delta_2]A_{\mu  } &=2i(\bar\epsilon_2\Gamma^\nu\epsilon_1) F_{\nu\mu  } \ ,\\
[\delta_1,\delta_2]\Psi &=2i(\bar\epsilon_2\Gamma^\nu\epsilon_1) {\cal D}_\nu \Psi +\dots\ ,
\end{aligned}
\end{equation}
 where the ellipses denote terms which vanish on-shell.

The second condition in (\ref{econ-D-branes}) tells us how much supersymmetry is preserved by the defect.
Note that $i\omega_{IJ}$ is a Hermitian and anti-symmetric matrix:
therefore it is diagonalisable and its eigenvalues come in pairs differing by a sign.
Thus we can introduce orthogonal complex coordinates $Z^A$ such that
\begin{equation}
  \omega\indices{^A_B} = \pmqty{\dmat{\varepsilon_1, \varepsilon_2, \varepsilon_3, \varepsilon_4}},
\end{equation}
and similarly for the complex conjugates.
For generic choices of $\varepsilon$'s, $\omega$ breaks the R--symmetry group from $SO(9-p)$ to a product of $U(1)$'s.
The deformation induces a twisting of the gauge theory connection, $D_\mu  \to {\cal D}_\mu$ (Eq.~\eqref{defD}), that includes a common extra $U(1)$, under which
each of the complex scalar fields can be thought of as carrying charge $\varepsilon_A$.
Depending on the choice of these charges, the final configuration preserves between \(0\) and \(8\) Killing spinors (unbroken supersymmetries).
One last comment is needed about chirality.
In the \D1~brane case (\emph{i.e.} for a two-dimensional gauge theory), some of the configurations of the \(\varepsilon\) in the background preserve a chiral (with respect to the operator \(\Gamma^{01}\)) subset of the supersymmetries.
These are inherited by the theory on the brane that can then be chiral with \((4,0)\) or \((6,0)\) supersymmetry (see Table~\ref{tab:D1-SUSY}).
It is easy to verify that no chiral configurations are possible for higher-dimensional theories.

\begin{table}
  \centering
  \begin{tabular}{lrr}
    \toprule
    conditions on $\varepsilon_{A}$ & unbroken SUSYs & chirality (w.r.t. $\Gamma^{01}$) \\
    \midrule
    $\sum_{A=1}^{4}\varepsilon_{A} = 0$                                                & $0$            & -                                \\[5pt]
    $\sum_{A=1}^{3}\varepsilon_{A} = 0\,\&\, \varepsilon_{4}=0$                     & $4$            & $(2,2)$                          \\[5pt]
    $\varepsilon_{1}=\pm\varepsilon_{2}\, \&\, \varepsilon_{3} = \varepsilon_{4}=0$ & $8$            & $(4,4)$                          \\[5pt]
    $\varepsilon_{1} = \varepsilon_{2}\,\&\, \varepsilon_{3} = \varepsilon_{4}$     & $4$            & $(4,0)$                          \\[5pt]
    $\varepsilon_{1} = \varepsilon_{2}=\varepsilon_{3}=\varepsilon_{4}$                & $6$            & $(6,0)$                          \\[5pt]
    \bottomrule
  \end{tabular}
  \caption{Unbroken supersymmetries in the \D1 action for different choices of the \(\varepsilon\) parameters.}
  \label{tab:D1-SUSY}
\end{table}

\subsection{Wilson lines of global symmetries}
\label{sec:wilson-lines}

We can consider a general supersymmetric gauge theory that has a Lagrangian  $\mathcal{L}$ and global symmetry $H$. By standard techniques we can gauge this symmetry by introducing an additional gauge field $\mathcal{B}_\mu$ which takes values in ${\rm Lie}(H)$ and modifying the covariant derivative to
\begin{align}
\mathcal{D}_\mu = D_\mu - \mathcal{B}_\mu\ .
\end{align}
The new action is obtained from the old by the replacement $D_\mu\to \mathcal{D}_\mu$:
\begin{align}
  \mathcal{L}_{\text{deformed}}=   \mathcal{L}(D_\mu\to \mathcal{D}_\mu) \ ,
\end{align}
where $\mathcal{G}_{\mu\nu}$ is the field strength of $\mathcal{B}_\mu$. If $\mathcal{B}_\mu$  is a flat connection then the effect of this change is locally trivial. So therefore the supersymmetry variation of this action must be of the form, assuming $\delta \mathcal{ B}_\mu=0$,
\begin{equation}\label{SOmega}
\delta \mathcal{L}_{\text{deformed}} =   i\bar\epsilon \Tr(\mathcal{ G}_{\mu\nu}\Omega^{\mu\nu})\ ,
\end{equation}
where $\Omega_{\mu\nu}$ is some expression in the original fields.  We can fix this by including a new Lagrange multiplier field $\chi^{\mu\nu}$ into the action
\begin{equation}
\mathcal{L}_{\text{deformed}} =   \mathcal{L}(D_\mu\to \mathcal{D}_\mu) - i\Tr(\chi^{\mu\nu}\mathcal{ G}_{\mu\nu}) \   ,
\end{equation}
and set
\begin{equation}
\delta \chi^{\mu\nu}  =  i\bar\epsilon \Omega^{\mu\nu}\ .
\end{equation}
Thus $\delta \mathcal{L}_{\text{deformed}}=0$ and we have a new supersymmetric gauge theory with $H$ local.

However
the gauge field $\mathcal{B}_\mu$ carries no degrees of freedom and is constrained to be flat.\footnote{One could also add a kinetic term 
$-\frac{1}{4g'^2}\Tr(\mathcal{G}^2)$ with some coupling $g'$ so as to make $\mathcal{B}_\mu$ dynamical, this would preserve supersymmetry provided $\mathcal{B}_\mu$ is taken to be a supersymmetry singlet.}   Nevertheless this still allows us to introduce a Wilson line for $\mathcal{B}_\mu$:
\begin{equation}
\mathcal{B}_\mu = \xi_\mu \omega\ ,
\end{equation}
where $\xi_\mu$ is constant and $\omega$ is an element of $Lie(H)$. Indeed  we see from (\ref{SOmega}) that $\delta S_{\text{deformed}}=0$ if $\mathcal{B}_\mu$ is flat and therefore we do not need to introduce $\chi^{\mu\nu}$ to preserve supersymmetry. Thus the story we have told clearly generalises to include additional matter content and, since it is simply based on weak-gauging, it should also apply to non-Lagrangian theories.

\section{Lift to M-theory}
\label{sec:lift-M}
The fluxtrap background in Eq.~(\ref{eq:fluxtrap}) can be lifted to M-theory and is given by
\begin{equation}
  \begin{aligned}
    \dd s^2_{11} &=  \Delta^{1/6} \left[ - (\dd x^0)^2 + \left(\delta_{IJ}- \frac {U_I U_J}{\Delta^2}\right)  \dd x^I  \dd x^J   + \frac{(\dd x^1)^2+(\dd x^{10})^2}{\Delta^2} \right] ,\\
    C_3 &= \frac{1}{\Delta^2} \dd x^1 \wedge dx^{10}\wedge U.
  \end{aligned}
\end{equation}
We want to study \M2~branes extended along $x^0,x^1,x^{10}$ in this configuration.
The analysis is clearly similar to the \D{p}~brane story above.
Let us first consider the  \ac{blg} model~\cite{Bagger:2006sk,Gustavsson:2007vu,Bagger:2007jr,Bagger:2007vi} of two \M2~branes. Here the undeformed action is
\begin{multline}
S_{BLG} =  -\int \frac{1}{2}\langle D_\mu X^I,  D^\mu X^I\rangle  +\frac{i}{2}\langle \bar\Psi,\Gamma^\mu D_\mu \Psi\rangle \\
+\frac{1}{4}\langle \bar\Psi,\Gamma^{IJ}[X^I,X^J,\Psi]\rangle+\frac16 \langle [X^I,X^J,X^K],[X^I,X^J,X^K]\rangle  - \mathcal{L}_{ CS} ,
\end{multline}
where $\langle\ \cdot\ ,\ \cdot\ \rangle$ is the inner-product on the 3-algebra, $[\ \cdot\ ,\ \cdot\ ,\ \cdot\ ]$ the totally anti-symmetric product (subject to the fundamental identity), $\mathcal{L}_{ CS} $ is a Chern--Simons term for $su(2)\oplus su(2)$ with opposite levels. The matter fields are in the bi-fundamental of $SU(2)\times SU(2)$ or $(SU(2)\times SU(2))/{\mathbb Z}_2$. This is invariant under the supersymmetry transformations
\begin{equation}
  \begin{aligned}
    \delta X^I & = -i\bar\epsilon \Gamma^I\Psi,\\
    \delta \tilde A_\mu (\ \cdot\ )& = -i\bar\epsilon \Gamma_\mu\Gamma^I [X^I,\Psi,\ \cdot\ ],\\
    \delta \Psi & = \Gamma^\mu\Gamma^I{ D}_\mu X^I   -\frac{1}{6}\Gamma^{IJ
      K}[X^I,X^J,X^K]\epsilon\ ,
  \end{aligned}
\end{equation}
where $\Gamma_{012}\epsilon=\epsilon$ and $\Gamma_{012}\Psi=-\Psi$.

The coupling to the background fluxes was discussed in~\cite{Lambert:2009qw}. The relevant term in this case is
\begin{equation}
S_{\text{WZ}} = \int \varepsilon^{\mu\nu\lambda}C_{\mu IJ}\langle D_\nu X^I , D_\lambda X^J\rangle\ .
\end{equation}
This has the same  effect as before leading to a deformed action $S_{BLG}+S_{\text{WZ}}+S_{2}$ which again is obtained by the replacement (\ref{defD}) applied to the \ac{blg} model. The preserved supersymmetries then satisfy
\begin{align}
  \Gamma_{01\widehat{10}}\epsilon &=\epsilon\ ,& \omega_{IJ}\Gamma^{IJ}\epsilon &=0\ .
\end{align}
As with the \D{p}~branes one finds that the supersymmetry algebra closes on-shell to translations, gauge transformations and R-symmetry. 
 
For the \ac{abjm} theory of \M2~branes the story is slightly more complicated.
Starting with the \ac{abjm}/\ac{abj} models~\cite{Aharony:2008ug,Aharony:2008gk}, the undeformed  action is (here we used the conventions of~\cite{Bagger:2008se}, but see also~\cite{Gaiotto:2008cg,Grignani:2008is})
\begin{equation}
  \label{niceaction}
  \begin{aligned}
     S ={}& -\Tr\int D^\mu Z_A D_\mu Z^A +
    i \bar\psi^A \gamma^\mu D_\mu\psi_A  +V-\mathcal{L}_{CS}\\
    & +i \bar\psi^A [\psi_{A},
    Z^B;Z_{B}] -2i \bar\psi^A[\psi_{B},Z^B;Z_{A}]\\
    &-\frac{i}{2}\varepsilon_{ABCD} \bar\psi^A[
    Z^C,Z^D;\psi^B] +\frac{i}{2}\varepsilon^{ABCD} Z_D [\bar
    \psi_{A},\psi_B; Z_{C}]\ ,
  \end{aligned}
\end{equation}
where
\begin{equation}
[Z^A,Z^B; Z_C] = \lambda (Z^AZ_CZ^B-Z^BZ_CZ^A)\ .
\end{equation}
The supersymmetry transformations are
\begin{equation}
  \label{finalsusy}
  \begin{aligned}
    \delta Z^A &=  i\bar\epsilon^{AB}\psi_{B } ,\\
    \delta \psi_{B } &= \gamma^\mu D_\mu Z^A \epsilon_{AB} + [Z^C,
    Z^A; Z_{C } ]\epsilon_{AB}+
    [Z^C, Z^D;Z_{B}]\epsilon_{CD}, \\
    \delta \tilde A_\mu (\ \cdot \ ) &= -i\bar\epsilon_{AB}\gamma_\mu
    [\ \cdot\ ,Z^A ;\psi^B] + i\bar\epsilon^{AB}\gamma_\mu [\ \cdot\
    ,\psi_B;Z_A] .
  \end{aligned}
\end{equation}
The relevant flux term is now~\cite{Lambert:2009qw}
\begin{equation}
  S_{\text{WZ}} = \int \varepsilon^{\mu\nu\lambda} C\indices{_\mu^A_B} \Tr( D_\nu Z_A D_\lambda Z^B) +  \varepsilon^{\mu\nu\lambda} C\indices{_\mu_A^B} \Tr( D_\nu Z^A D_\lambda Z_B) .
\end{equation} 
It is important to note that a general  $\omega_{AB}$ defines an element of $\mathfrak{su}(4)\oplus \mathfrak{u}(1)$. However only $\mathfrak{su}(4)$ generates an R--symmetry $SU(4)$. The remaining $\mathfrak{u}(1)$ generates a $U(1)$ group which is gauged in the \ac{abjm}/\ac{abj} models. Furthermore while $Z^A$ and $\psi^A$ transform in the same representation of  $\mathfrak{su}(4)$ they carry opposite $\mathfrak{u}(1)$ charges. Therefore it is useful to write
\begin{align}
  \omega\indices{_A^B} & = i\omega_0 \delta\indices{_A^B} + \tilde \omega\indices{_A^B} \ , &  \tilde \omega\indices{_A^A}=0\ .\
\end{align}
Thus a general flux induces two different currents in the worldvolume theory:
\begin{equation}
S_{\text{WZ}}  = \int \xi^\lambda i\omega_0 J_\lambda  + \xi^\lambda \tilde \omega_A{}^B J_\lambda^{A}{}_B,
\end{equation}
where
\begin{align}
  j_\mu   & = \Tr(Z^AD_\mu Z_A-D^\mu Z^AZ_A)-i\Tr(\bar\psi^AD_\mu \psi_A) , \\
  j_\mu^A{}_B & = \Tr(Z^AD_\mu Z_B-D^\mu Z^AZ_B)+i\Tr(\bar\psi^AD_\mu \psi_B)\ ,
\end{align}
are the $\mathfrak{u}(1)$ gauge and $\mathfrak{su}(4)$ R--symmetry currents respectively.
With these points noted, one again finds that the effect of the flux is to induce a connection taking values in the  R--symmetry and $U(1)$ Lie algebras:
\begin{align}
  D_\mu Z^A & \to \mathcal{D}_\mu Z^A = D_\mu Z^A - i\xi_\mu\omega_0 Z^A- \xi_\mu\tilde \omega_B{}^A Z^B ,\\
  D_\mu \psi^A & \to \mathcal{D}_\mu \psi^A = D_\mu \psi^A + i\xi_\mu\omega_0\psi^A - \xi_\mu\tilde \omega_B{}^A \psi^B .
\end{align}
The preserved supersymmetries satisfy
\begin{equation}\label{presusy}
  \tilde \omega\indices{_A^C} \epsilon_{CB} = \tilde \omega\indices{_B^C}\epsilon_{CA}\ .
\end{equation}
As before one finds that the supersymmetry algebra closes on-shell to translations, gauge transformations and R-symmetry.

To examine this condition we choose a coordinate system where
\begin{equation}
  \omega\indices{_A^B} = i \pmqty{\dmat{\varepsilon_1, \varepsilon_2, \varepsilon_3, \varepsilon_4}}\ ,
\end{equation}
thus
\begin{small}
\begin{align}
  \omega_0                    & = \frac{1}{4} \pqty{\varepsilon_1 + \varepsilon_2 + \varepsilon_3 + \varepsilon_4}   , \\
  \tilde \omega\indices{_A^B} & = \pmqty{\dmat{3\varepsilon_1-\varepsilon_2-\varepsilon_3-\varepsilon_4, 3\varepsilon_2-\varepsilon_1-\varepsilon_3-\varepsilon_4 , 3\varepsilon_3-\varepsilon_1-\varepsilon_2-\varepsilon_4 , 3\varepsilon_4-\varepsilon_1-\varepsilon_2-\varepsilon_3}}.
\end{align}
\end{small}
The condition (\ref{presusy}) is then simply that $\epsilon_{AB}$ is preserved if and only if
 \begin{equation}\label{econ}
 \varepsilon_A+\varepsilon_B = \varepsilon_C+\varepsilon_D\ ,
 \end{equation}
where $A,B,C,D$ are all distinct. Thus if $\epsilon_{AB}$ is preserved then so is $\epsilon_{CD}$ with $C,D\ne A,B$. Since each $\epsilon_{AB}$ has two real spinor components, the total number of preserved supersymmetries is  a multiple of 4. In particular, for a generic choice, there are no supersymmetries.   If (\ref{econ}) is satisfied for any choice of pairs of $\varepsilon_A$'s, then there are $4$ supersymmetries. If in addition, there are two equal $\varepsilon_A$'s then there are $8$ supersymmetries. 
If all $\varepsilon_A$ are equal, then there are $12$ preserved supersymmetries. This last choice corresponds to $\omega\indices{_A^B} = i\omega_0 \delta\indices{_A^B}$ and the resulting deformation simply adds a Wilson line to the $A^L_\mu-A^R_\mu$ $U(1)$ gauge field, without any additional twisting with the normal bundle,  and does not break any supersymmetries of the \ac{abjm}/\ac{abj} model. Therefore one finds a one-dimensional maximally supersymmetric family of deformed \ac{abjm} models.

\section{Higher-dimensional deformations}
\label{sec:higher-dimensional-defects}

In the previous section we constructed relatively simple deformations of gauge theories that correspond to twisting the covariant derivative with   the R-symmetry. From the String Theory point of view these deformations arise  from  a brane in a flux background (the same flux that couples to the brane) but which has been mis-aligned in the sense the two legs of the flux lie off the brane. These deformations can also be interpreted as arising from the presence of 1-dimensional Wilson line defect. As such these deformations are rather generic and we explicitly constructed it for \D{p}~branes and M2~branes.

We now want to consider a related deformation which also arises from putting branes in a flux background  (and again the same flux that couples to the brane) which is mis-aligned. However these deformations rely on the fact that the bulk fluxes are always self-dual in the sense that if there is a non-vanishing flux arising from $C_{p+1}$ then there must also be a non-vanishing $C_{7-p}$. In the examples of the previous section the higher-form flux only couples to the brane through higher-derivative terms and, at low energy, can be neglected and decoupled. In this section we explore examples where both fluxes couple equally to the brane.
 
The simplest example of such a flux is to consider a 4-form $C_4$ in type IIB String Theory whose field strength $\dd C_4$ must be self-dual. Such a flux can be obtained from the fluxtrap solution above by setting $\varepsilon_1=\varepsilon_2=0$ and T--dualizing twice. To first order the resulting flux is $\dd C_4 = \Xi\wedge \omega$ where
\begin{equation}\label{eq:Unew}
 \omega  = -\varepsilon  \dd x^6\wedge \dd x^7   \pm \varepsilon \dd x^8\wedge  
 \dd x^9 
\end{equation}
and $\Xi = \pm \star_6 \Xi$ where $\star_6$ is the Hodge dual in the $x^0,...,x^5$ plane. 

In this section we first consider a \D5~brane placed  along $x^0,...,x^5$ in this \ac{rr} four-form background, and then focus on a \D3~brane. In the former case, we will find a deformation that appears to arise from two intersecting 3-dimensional defects while in the latter from two  2-dimensional ones. We will then describe how to reduce the gauge theory via T-dualities to a variety of lower-dimensional theories, which generally carry either 2- or 3-dimensional defects.

\subsection{The D5~brane}
\label{sec:D5}
\paragraph{RR four-form background I. }
Let us describe the ten-dimensional background of our interest first. It can be derived starting from a flat background with Melvin identifications (see Appendix~\ref{sec:Melvin} for the derivation).
This background contains a non-zero \ac{rr} four-form, but no dilaton or Kalb--Ramond field:
\begin{equation}
  \label{eq:bg-D5}
  \begin{aligned}
    g_{mn}\dd{x^{m}}\dd{x^{n}} &= \Delta \eta_{\alpha\beta} \dd{x^{\alpha}} \dd{x^{\beta}} + \frac{\delta_{ab} \dd{x^{a}}\dd{x^{b}}}{\Delta}+  \left(\Delta \delta_{IJ}    - \frac{U_I U_J}{\Delta}\right) \dd{x^{I}} \dd{x^{J}} \\
    C_{4} &= U \wedge \left(-\dd x^0\wedge \dd x^1\wedge \dd x^5  + \frac{\dd x^2\wedge \dd x^3\wedge \dd x^4}{\Delta^{2}}\right),
  \end{aligned}
\end{equation}
where \(\alpha, \beta = 0,1,5\); \(a,b = 2,3,4\); \(I, J = 6,7,8,9\) and $\Delta$ as in Eq.~\eqref{eq:Delta}. The deformation parameter $\omega_{IJ}$ is given by $\omega_{76}=\pm\omega_{89} = 2\varepsilon$.
It is convenient to recast the four-form potential in the form
\begin{equation}
  C_{4} = -\frac{1}{3!}\omega_{IJ}x^{J} \dd{x^{I}} \wedge \left(\Xi_{\alpha\beta\gamma}  \dd x^\alpha\wedge \dd x^\beta\wedge \dd x^\gamma + \Xi_{abc}\frac{\dd x^a\wedge \dd x^b\wedge \dd x^c}{\Delta^{2}}\right),
\end{equation}
where \(\Xi\) is an anti-self dual tensor
\begin{align}
  \omega_{IJ} &= \mp\frac{1}{2}\epsilon_{IJKL}\omega^{KL}\,, & \Xi_{\mu\nu\rho} &= \mp \frac{1}{3!}\epsilon_{\mu\mu\rho\mu' \nu' \rho'}\Xi^{\mu' \nu' \rho'}\,.
\end{align}
For convenience we will generally take
\begin{align}
  \label{eq:xi-D5}
  \Xi &= \frac{1}{2\cdot 3!}\left(\Xi_{\alpha\beta\gamma} \dd x^\alpha\wedge \dd x^\beta\wedge \dd x^\gamma+\Xi_{abc}\dd x^a\wedge \dd x^b\wedge \dd x^c\right) \nonumber\\
  &= \frac{1}{2} \pqty{-\dd x^0\wedge \dd x^1\wedge \dd x^5  \pm \dd x^2\wedge \dd x^3\wedge \dd x^4}\,,
\end{align}
which will have the effect of splitting the D~brane worldvolume into two subspaces $\{x^{\alpha}\}_{\alpha=0,1,5}$ and $\{x^{a}\}_{a=2,3,4}$.

\paragraph{Supersymmetric D5~brane action.}

Let us place a probe \D5~brane in the static embedding along  $\{x^\mu\}_{\mu=0,...,5}$ and compute its effective action for bosons and fermions. 
We start with the \ac{cs} term. On the six-dimensional world-volume, only the product of the \ac{rr} four-form and the gauge field strength contribute:
\begin{equation}
  S_{\text{CS}} = \frac{1}{g^{2}}\int \hat C_4 \wedge F_2 = - \frac{1}{2g^{2}}\int \dd[6]{x} \pqty{\frac{1}{\Delta^{2}} \Xi_{\alpha \beta \gamma} F^{\beta \gamma} \omega_{IJ} X^{J}\partial^{\alpha}X^{I} + \Xi_{a b c } F^{b c} \omega_{IJ}X^{J}\partial^{a}X^{I}} .
\end{equation}
This form suggests, in analogy with the Wilson line example of Sec.~\ref{sec:line-defects}, to introduce a connection
\begin{equation}
  \label{eq:connection-D5}
  \mathcal{A}_{\mu} = \frac{1}{2} \Xi_{\mu \nu \rho} F^{\nu \rho}\ ,
\end{equation}
and a covariant derivative \(\covD\) that acts on the scalars as
\begin{equation}
\label{eq:covD-real}
  \covD_{\mu}X^{I} = \partial_{\mu}X^{I}+\mathcal{A}_{\mu}\omega^{IJ}X_{J}\,.
\end{equation}
This can be seen as resulting from a non-minimal coupling of the scalars to the gauge field.
In this way one can rewrite the bosonic part of the Abelian action, to quadratic order in the derivatives,  in the compact form
\begin{align}
  \label{eq:DBI-D5}
  S_{B}^{\D5} &= -\frac{1}{g^{2}}\int \dd[6]{x} \bqty{\frac{1}{2}\sqrt{-g} g^{\mu\nu}g_{IJ} \covD_{\mu}X^{I}\covD_{\nu}X^{J} +\frac{1}{4}\eta^{\mu\rho}\eta^{\nu\sigma} F_{\mu\nu}F_{\rho\sigma}}\,,
\end{align}
where $g_{IJ}$ and $g_{\mu\nu}$ refer to the  metric in the bulk \eqref{eq:bg-D5} pulled-back to the brane: 
\begin{equation} \label{gbrane}
  \begin{aligned}
    g_{\alpha\beta} &= \Delta \eta_{\alpha\beta} ,\\
    g_{ab} &= \Delta^{-1} \delta_{ab} ,\\
    g_{IJ} &= \Delta \delta_{IJ} - \frac{U_IU_J}{\Delta} ,\\
  \end{aligned}
\end{equation} 
and $\sqrt{-g}$ is the square-root of the worldvolume components of the metric (which is in fact \(1\)).
The latter breaks the Lorentz symmetry for the (1+5)-dimensional spacetime. 
The curved metric both in real and in field space is encoded in the kinetic term for the scalars.  
  When we restrict ourselves to the linearised deformation, only the \ac{cs} action contributes to $\mathcal{O}(\varepsilon)$.
\bigskip
\\
Next, we study the fermionic part. The covariant derivative \eqref{eq:covD-real} emerges also in the fermionic action, computed directly in terms of the pullback of the background fields (see~\cite{Martucci:2005rb}).
By choosing a canonical gauge-fixing condition for the 64-component spinor $\Theta = (\Psi, \tilde \Psi)^t$
\begin{equation}
  \pqty{\Gamma^{10} \otimes \sigma_{3}} \Theta = - \Theta   \Rightarrow \tilde \Psi=0\,,
\end{equation}
a tedious computation shows that the Dirac action for the \D5~brane is
\begin{equation}\label{eq:fermionicD5}
  \begin{aligned}
    S^{\D5}_{F}  =\frac{i}{2g^{2}}\int \dd[6]{x} \bar{\Psi} \bqty{\Gamma^{\mu}\covD_{\mu} + \frac{1}{4}\Xi_{\mu\nu\rho}\Gamma^{\nu\rho}\omega_{IJ}\partial^{\mu}X^{I}\Gamma^{J}} \Psi + \mathcal{O}(\varepsilon^{2}),
  \end{aligned}
\end{equation}
where the covariant derivative acts on the spinors as
\begin{equation}
  \covD_{\mu}\Psi = \partial_{\mu}\Psi + \frac{1}{4} \mathcal{A}_{\mu} \omega_{IJ}\Gamma^{IJ} \Psi = \partial_{\mu}\Psi + \frac{1}{8} \Xi_{\mu \nu \rho} F^{\nu \rho} \omega_{IJ}\Gamma^{IJ}\Psi \,.
\end{equation}

\paragraph{Spacetime and R-symmetries.}

Let us consider the symmetries in the action.  The insertion of a \D5~brane breaks the Lorentz group from \(SO(1,9)\) to \(SO(1,5) \times SO(4)\). The former is a six-dimensional Lorentz symmetry whereas the latter is an R-symmetry for four transverse scalars.
However, the flux as well as the metric in the deformed background further break \(SO(1,5)\) to \(SO(1,2) \times SO(3)\). It is natural to interpret this symmetry breaking as the consequence of the presence of two defects in the six-dimensional theory, which are extended in $\{x^{\alpha}\}_{\alpha=0,1,5}$ and $\{x^{a}\}_{a=2,3,4}$.  

Next, we discuss the R-symmetry breaking.
By turning on the fluxes, the real scalars are twisted in the covariant derivatives, and this makes the remaining R-symmetry hard to see in the \ac{dbi} action.
The way out is to rewrite the action \eqref{eq:DBI-D5} using a doublet  of complex scalars defined as 
\begin{equation}
\label{eq:doublet-D5}
\mathbf{Z} =\left( \begin{matrix}
 X^{6} + iX^{7} \\  X^{8}\mp iX^{9}
 \end{matrix}
\right) , 
\end{equation}
where the sign difference between $X^{7}$ and $X^{9}$ reflects $\varepsilon_{1} = \mp\varepsilon_{2}$.
Then, Eq.~\eqref{eq:DBI-D5} takes the form
\begin{align}
\label{eq:DBI-D5-comp}
S^{\D5}_{B} &= -\frac{1}{g^{2}}\int d^{6}x \frac{1}{2}\sqrt{-g} g^{\mu\nu} \biggr[\Delta (\mathcal{D}_{\mu}\mathbf{Z})^{\dagger}\, \mathcal{D}_{\nu}\mathbf{Z} - \frac{\mathcal{J}_{\mu}\mathcal{J}_{\nu}}{\Delta}\biggl]+\frac{1}{4}\eta^{\mu\rho}\eta^{\nu\sigma}F_{\mu\nu}F_{\rho\sigma}\,,
\end{align}
where the $\dagger$ denotes a Hermitian conjugation and
\begin{align}
\label{eq:Current}
\mathcal{J}_{\mu} &= U_{I}\mathcal{D}_{\mu}X^{I} =\frac{i\varepsilon}{2}\left(\mathbf{Z}^{\dagger}\mathcal{D}_{\mu}\mathbf{Z} - (\mathcal{D}_{\mu}\mathbf{Z})^{\dagger}\mathbf{Z}\right),
\end{align}
with a covariant derivative defined by 
\begin{equation}
\label{eq:cov2-D5}
\mathcal{D}_{\mu}\mathbf{Z} = \partial_{\mu}\mathbf{Z} + 2i\varepsilon \mathcal{A}_{\mu}\mathbf{Z}\,.
\end{equation} 
In this form the R-symmetry is manifestly broken to \(SU(2)\), under which the doublet $\mathbf{Z}$ transforms in the fundamental representation.  

\paragraph{Equations of motion.}
Since our D5~brane action preserves Lorentz invariance only in \(1 + 2 \) dimensions, we can also think of the preserved supersymmetries as realising a \(d = 3 \), \(\mathcal{N} = 4\) superalgebra, if the gauge theory is dimensionally reduced on the \(T^3\) generated by $\{x^{a}\}_{a=2,3,4}$. Thus, it will be instructive to analyse our action by ignoring the dependence of the fields on $\{x^{a}\}$. This will allow us to see features of three-dimensional theories, such as a dual photon.

First, let us derive the equations of motion for the scalars focusing on the bosonic action. Note that the $(\alpha,\beta)$-component in the kinetic term of 	\eqref{eq:DBI-D5-comp} is canonicalised whereas the $(a,b)$-component is not. This implies that it is no longer possible to deal with both directions on equal grounds, which results from the presence of defects.

Varying the action with respect to $\mathbf{Z}^{\dagger}$, one finds for $\mathbf{Z}$
\begin{multline}
  (\mathcal{D}_{\mu} \Delta g^{\mu\nu} \mathcal{D}_{\nu})\mathbf{Z} + i\varepsilon\left(\mathcal{D}_{\mu}\left(\frac{g^{\mu\nu}}{\Delta}\mathcal{J}_{\nu}\mathbf{Z}\right) + \frac{g^{\mu\nu}}{\Delta}\mathcal{J}_{\mu}\mathcal{D}_{\nu}\mathbf{Z} \right)\\
  -\varepsilon^{2} \frac{\eta^{\alpha\beta}\mathcal{J}_{\alpha}\mathcal{J}_{\beta}}{\Delta^{4}}\mathbf{Z} - \eta^{ab}(\mathcal{D}_{a}\mathbf{Z})^{\dagger}(\mathcal{D}_{b}\mathbf{Z})\mathbf{Z} =0\,.
\end{multline}
In the second line, both the indices $\alpha,\beta$ and $a,b$ appear separately.  

Second, we derive the equation of motion for the gauge fields. Unlike for the scalar fields, one can obtain two simpler equations:
\begin{align}
\label{eq:6dgaugeEOM}
&\partial^{\alpha}\left( F_{\alpha\beta}-\frac{2}{\Delta^{2}}\Xi_{\alpha\beta\gamma}\mathcal{J}^{\gamma}\right)+\partial^{a}F_{a\beta} = 0\,,\\
&\partial^{a}\left( F_{ab}-2\Xi_{abc}\mathcal{J}^{c}\right)+\partial^{\alpha}F_{\alpha b} = 0\,,
\end{align}
where $\mathcal{J}_{\mu}$ is given by \eqref{eq:Current}.
Note that the mixed components $F_{a\alpha}$ do not receive any corrections from the deformation. One could regard the effect of the deformation as shifting the field strengths $F_{\alpha\beta}$ and $F_{ab}$ by $\mathcal{J}_{\mu}$. 

As earlier discussed for the dimensional reduction to three dimensions, it is natural to restrict the spacetime dependence of the fields and to analyse the equations of motion. We expect a \(1+2\) dimensional sector $\{x^{\alpha}\}_{\alpha=0,1,5}$ to be coupled inside the six-dimensional worldvolume. For a simple interpretation, suppose that every field depends exclusively on the $\{x^{\alpha}\}$-plane.
Then the second term in \eqref{eq:6dgaugeEOM} drops and one finds that the remaining \ac{eom} describes the conservation of a current
\begin{equation}
  \del^\alpha \pqty{ F_{\alpha \beta} - \frac{2}{\Delta^2} \Xi_{\alpha \beta \gamma} \mathcal{J}^{\gamma}} = 0 .
\end{equation}
The system is now effectively three-dimensional so it is convenient to rewrite the \ac{eom} as
\begin{equation}
  \dd \star_{3} \pqty{ F - \frac{1}{\Delta^2} \star_{3} \mathcal{J}} = \dd \pqty{ \star_{3}F + \frac{1}{\Delta^{2}} \mathcal{J}} = 0 ,
\end{equation}
where we have observed that the restriction of \(\Xi\) to the three-dimensional subspace is simply the Hodge star.
The equation admits the solution
\begin{equation}
  \star_{3} F + \frac{1}{\Delta^{2}} \mathcal{J} = \dd \phi ,
\end{equation}
which we can understand in terms of a dual scalar living in the deformed theory,
\begin{equation}\label{eq:dualscalarD5}
\partial_{\alpha} \phi \equiv 2\mathcal{A}_{\alpha} + \frac{\mathcal{J}_{\alpha}}{\Delta^{2}}\,.
\end{equation}

\paragraph{Supersymmetry.}
By construction, the ten-dimensional background in Eq.~\eqref{eq:bg-D5} preserves \(16\) Killing spinors since it is related to Melvin space via a series of dualities. In addition, the D5~brane on a classical configuration normally breaks half of the supersymmetry. Therefore the \D5~brane embedded in Eq.~\eqref{eq:bg-D5} is expected to preserve \(8\) supercharges, \emph{i.e.} it can be seen as a maximally supersymmetric six-dimensional theory in the presence of a half-\ac{bps} defect.

We have shown that the deformation changes the form of the action, as discussed in the previous section. In this section we also present how the supersymmetry variations are modified. These turn out to be relatively complicated and 
instead of trying to construct these transformations directly by making an ansatz  it is more convenient and straightforward to derive them from String Theory based on the results of~\cite{Martucci:2005rb}. The following analysis is restricted to the first order $\mathcal{O}(\varepsilon)$. The action is given by
\begin{equation}\label{Sa}
\begin{aligned}
S &= -\frac{1}{g^2}\int d^6 x \frac14 F_{\mu\nu}F^{\mu\nu} + \frac{1}2 {\cal D}_\mu X^I{\cal D}^\mu X^I - \frac{i}{2}\bar\Psi\Gamma^\mu{\cal D}_\mu\Psi -\frac{i}{8}\bar\Psi{\cal D}^\mu X^I\Gamma^J\omega_{IJ}\Xi_{\mu\nu\lambda}\Gamma^{\nu\lambda}\Psi ,\\
&= -\frac{1}{g^2}\int d^6 x \frac14 F_{\mu\nu}F^{\mu\nu} + \frac12 \partial_\mu X^I\partial^\mu X^I - \frac{i}{2}\bar\Psi\Gamma^\mu\partial_\mu \Psi\\
& \qquad + \frac 12 \Xi_{\mu\nu\lambda}F^{\nu\lambda}\omega^{IJ}X^J\partial^\mu X^I - \frac{i}{16}\bar\Psi\Gamma^\mu\Xi_{\mu\nu\lambda}F^{\nu\lambda}\omega_{IJ}\Gamma^{IJ} \Psi-\frac{i}{8}\bar\Psi\partial^\mu X^I\Gamma^J\omega_{IJ}\Xi_{\mu\nu\lambda}\Gamma^{\nu\lambda}\Psi .\\
\end{aligned}
\end{equation}

The first ingredient that we need is the gravitino supersymmetry variation  in type IIB supergravity:
\begin{align}
&\biggl[\nabla_m + \frac{1}{4\cdot2!}H_{mnp}\Gamma^{np}\sigma_{3}\nonumber\\
&\hspace{0.2in} + \frac{e^{\Phi}}{8}\left(F_{n}\Gamma^{n} (i\sigma_{2}) + \frac{1}{3!}F_{npq}\Gamma^{npq}\sigma_{1}+\frac{1}{2\cdot 5! }F_{npqrt}\Gamma^{npqrt}(i\sigma_{2})\right)\Gamma_{m} \biggr]{\cal E}(x) = 0\,.
\end{align} Since the background metric is flat and only the five-form flux contributes to the first order $\mathcal{O}(\varepsilon)$, the gravitino equation reads 
\begin{equation}
  \del_{m} {\cal E}(x) = \pqty{\frac{1}{4   \times 4!} \omega_{IJ}\Gamma^{IJ} \Xi_{\mu\nu\rho}\Gamma^{\mu\nu\rho}\Gamma_{m} \otimes (i\sigma_{2}) } {\cal E}(x)\ ,
\end{equation}
which is solved by
\begin{equation}
  {\cal E}(x) = \left( 1 + \frac{1}{4 \times 4!} \omega_{IJ} \Gamma^{IJ} \Xi_{\mu\nu\rho} \Gamma^{\mu\nu\rho} x^{m} \Gamma_{m} \otimes (i\sigma_{2}) \right) {\cal E}_0 \,,
\end{equation}
where \({\cal E}_0 = (\epsilon_0, \tilde \epsilon_0)^t\) is a doublet of ten-dimensional constant Majorana--Weyl spinors both preserved by the Melvin deformation. The fact that the supersymmetry is preserved by a rigid \D5~brane extended in $\{x^{\mu}\}_{\mu=0,...,5}$ translates into
\begin{equation}
\label{eq:Killing1}
  {\cal E}_0 =   \begin{pmatrix}
    \epsilon_0 \\ \tilde \epsilon_0
  \end{pmatrix} =
  \begin{pmatrix}
    \epsilon_0 \\ \Gamma_{012345} \epsilon_0
  \end{pmatrix}
\end{equation}
at the zero-th order of $\varepsilon$ and 
\begin{equation}
  \pqty{\omega_{IJ}\Gamma^{IJ}\otimes \mathbb{1} } {\cal E}_0 = 0\quad \Leftrightarrow\quad \omega_{IJ}\Gamma^{IJ}\epsilon_{0} =0\ ,
\end{equation}
at the first order $\order{\varepsilon}$.
As a result, only \(8\) free real parameters are left on the D5~brane worldvolume as expected.

The pullback of the Killing spinor on the \D5~brane is then
\begin{equation}
\label{eq:Killingspinor}
  \epsilon(X) = \bqty{1 + \frac{1}{4!}\Xi_{\mu\nu\rho}\Gamma^{\mu\nu\rho}\omega_{IJ}X^{I}\Gamma^{J}} \epsilon_0,
\end{equation}
where both \(\epsilon(X)\) and \(\epsilon_0\) are \(32\)-component spinors.
This leads to the following transformation rules up to order $\mathcal{O}(\varepsilon)$:
\begin{equation}
\begin{aligned}\label{Susya}
\delta X^I & = i\bar\epsilon(X)\Gamma^I\Psi  \\
&=i\bar\epsilon \Gamma^I\Psi  +  \frac{1}{4!}\bar\epsilon(\Xi\cdot \Gamma )\Gamma^K\omega^{JK}X^J \Gamma^I\Psi, \\
\delta A_\mu & = i\bar\epsilon(X)\Gamma_\mu\Psi  \\
&=i\bar\epsilon \Gamma_\mu\Psi +  \frac{1}{4!}\bar\epsilon(\Xi\cdot \Gamma )\Gamma^K\omega^{JK}X^J \Gamma_\mu\Psi , \\
\delta \Psi &= \frac12 \Gamma^{\mu\nu}F_{\mu\nu}\epsilon(X) + \Gamma^\mu\Gamma^I\partial_\mu X^I\epsilon(X)\\ 
&=\frac12 \Gamma^{\mu\nu}F_{\mu\nu  }\epsilon + \Gamma^\mu\Gamma^I{ D}_\mu X^I\epsilon  \\
&\qquad +\frac{1}{2\cdot 4!} \Gamma^{\mu\nu}F_{\mu\nu }(\Xi\cdot \Gamma )X^J\omega^{JK}\Gamma^K \epsilon+\frac{1}{4!} \Gamma^\mu\Gamma^I{D}_\mu X^I(\Xi\cdot \Gamma )X^J\omega^{JK}\Gamma^K \epsilon \ .
\end{aligned}
\end{equation}
At the first order $\mathcal{O}(\varepsilon)$, and with a choice of \(\kappa\)-symmetry gauge, the deformed supersymmetry transformations are completely captured by what in the ten-dimensional point of view is a non-constant supersymmetry parameter.
We are, however, in a decoupling limit without gravity and in six dimensions these are indeed rigid supersymmetry transformations but of higher order in the fields.
One can  check that these transformations leave the  action (\ref{Sa}) invariant to first order in the deformation. 

\paragraph{Non-Abelian generalisation.}

Let us examine the supersymmetry and action to first order for the non-Abelian theory. We saw that in the Abelian case the
 supersymmetry is corrected at first order due to the fact that the spacetime Killing spinor is no longer constant. Rather we found the Killing spinor, pulled-back to the worldvolume, to be
\begin{equation}
\epsilon(X) = \epsilon + \frac{1}{4!}(\Xi\cdot\Gamma )X^J\omega^{JK}\Gamma^K\epsilon\ ,
\end{equation}
plus higher order terms in both the deformation parameters and fermions. 
In the non-Abelian case we must take into account the ordering of the fields and include possible commutator terms. 

We begin by introducing generators $T^a$ of the Lie algebra such that 
\begin{equation}
{\rm Tr}(T^aT^b) = \delta^{ab}\ ,
\end{equation}
which we use as a metric that allows us to raise and lower Lie-algebra indices at will. A natural guess for the supersymmetry is that it corresponds to 
\begin{equation}
\begin{aligned}
\delta X^I_a & = i\bar\epsilon^b{}_a(X)\Gamma^I\Psi_b \\
\delta A_{\mu a} & = i\bar\epsilon^b{}_a(X)\Gamma_\mu\Psi_b \\
\delta \Psi_a &= \frac12 \Gamma^{\mu\nu}F_{\mu\nu b}\epsilon^b{}_a(X)+ \Gamma^\mu\Gamma^I D_\mu X^I_b\epsilon^b{}_a(X) -\frac{i}{2}\Gamma^{IJ}[X^I,X^J]_b\epsilon^b{}_a(X)\ , \\
\end{aligned}
\end{equation}
where 
\begin{equation}
\label{eq:nonAbe-Killingspinor}
\epsilon^a{}_b(X) = \delta^a_b\epsilon + \frac{1}{4!}(\Xi\cdot \Gamma )X^J_c\omega^{JK}\Gamma^Kd^{ac}{}_b\epsilon\ .
\end{equation}
Here   $d^{ac}{}_b$ is some invariant tensor.   Expanding out these expressions we find
\begin{equation}
\begin{aligned}
\delta X^I_a %
&=i\bar\epsilon \Gamma^I\Psi_a +  \frac{1}{4!}\bar\epsilon(\Xi\cdot \Gamma )\Gamma^K\omega^{JK}X^J_cd^{bc}{}_a\Gamma^I\Psi_b,\\
\delta A_{\mu a} %
&=i\bar\epsilon \Gamma_\mu\Psi_a +  \frac{1}{4!}\bar\epsilon(\Xi\cdot \Gamma )\Gamma^K\omega^{JK}X^J_cd^{bc}{}_a\Gamma_\mu\Psi_b, \\
\delta \Psi_a %
&=\frac12 \Gamma^{\mu\nu}F_{\mu\nu a}\epsilon + \Gamma^\mu\Gamma^I{ D}_\mu X^I_a\epsilon  -\frac{i}{2}\Gamma^{IJ}[X^I,X^J]_a\epsilon \\
&\qquad +\frac{1}{2\cdot 4!} \Gamma^{\mu\nu}F_{\mu\nu b}(\Xi\cdot \Gamma )X^J_c\omega^{JK}\Gamma^Kd^{bc}{}_a\epsilon+\frac{1}{4!} \Gamma^\mu\Gamma^I{D}_\mu X^I_b(\Xi\cdot \Gamma )X^J_c\omega^{JK}\Gamma^Kd^{bc}{}_a\epsilon \\
&\qquad-\frac{i}{2\cdot 4!}\Gamma^{KL}[X^K,X^L]_b(\Xi\cdot \Gamma )X^J_c\omega^{JI}\Gamma^Id^{bc}{}_a\epsilon\ .
\end{aligned}
\end{equation}
 We have checked that these variations close on the bosons (to lowest order in the fermions) so long as $d^{abc} = d^{cba}$. Thus we identify
\begin{equation}
d^{abc} = {\rm Str}(T^aT^bT^c) = \frac 12 ({\rm Tr}(T^aT^bT^c)+{\rm Tr}(T^cT^bT^a) )\ .
\end{equation}
However, in contrast to the previous case, the variations do not close on the R-symmetry.
Rather one finds
\begin{equation}
\begin{aligned}
[\delta_1,\delta_2]X^I_a &= v^\mu D_\mu X^I_a - i [\Lambda,X^I]_a \ ,\\
[\delta_1,\delta_2]A_{\mu a} &= v^\nu F_{\nu\mu a} + D_\mu \Lambda_a\ ,
\end{aligned}
\end{equation}
where $D_\mu$ is the undeformed covariant derivative and 
\begin{equation}
\begin{aligned}\label{vL}
v^\mu &=2i(\bar\epsilon_2\Gamma^\mu\epsilon_1),\\ 
\Lambda &= \frac{ 2i}{4!}(\bar\epsilon_2  (\Xi\cdot \Gamma) \Gamma^{JK}\epsilon_1 )\omega^{IJ}X^J_bX^Kd^{bc}{}_aT^a\ .
\end{aligned}
\end{equation}

Examining the closure on the fermions one finds that it includes terms involving 
$v^\mu$ which are not  translations $v^\mu D_\mu\Psi$ and cannot be  made to vanish by imposing an equation of motion. Presumably these can be cancelled by introducing ${\cal O}(\Psi^2\epsilon)$ into $\delta \Psi$. Such terms will not affect the closure of the bosons  or the invariance of the action at lowest order in the fermions and so we do not discuss them here.

To obtain the first order action we replace all the previous terms by the non-Abelian version and use the symmetrised trace prescription for the higher order terms:
 \begin{equation}
\begin{aligned}\label{Sna}
S &= -\frac{1}{g^2}{\rm Tr}\int d^6 x \frac14 F_{\mu\nu}F^{\mu\nu} + \frac12 D_\mu X^ID^\mu X^I \\
&\qquad\qquad\qquad- \frac{i}{2}\bar\Psi\Gamma^\mu D_\mu \Psi + \frac{1}{2}\bar\Psi\Gamma^I \comm{X^I}{\Psi}-\frac{1}{4} \comm{X^I}{X^J}\comm{X^I}{X^J}\\
& \qquad -\frac{1}{g^2}{\rm Str}\int d^6 x \frac 12 \Xi_{\mu\nu\lambda}F^{\nu\lambda}\omega^{IJ}X^J D^\mu X^I - \frac{i}{16}\bar\Psi\Gamma^\mu\Xi_{\mu\nu\lambda}F^{\nu\lambda}\omega_{IJ}\Gamma^{IJ} \Psi\\
&\qquad\qquad\qquad  -\frac{i}{8}\bar\Psi D^\mu X^I\Gamma^J\omega_{IJ}\Xi_{\mu\nu\lambda}\Gamma^{\nu\lambda}\Psi   +\frac{1}{2}\bar\Psi \Xi_{\mu\nu\lambda} \Gamma^{\mu\nu\lambda} \omega^{IJ}[X^I,X^J]\Psi\ .
\end{aligned}
\end{equation} 
Here we have included the last term which vanishes in the Abelian limit. We guessed its existence from looking at the on-shell conditions that arise from the incomplete closure of the fermions.
We have verified that (\ref{Sna}) is invariant under the supersymmetry up to first order in the deformation (and lowest order in the fermions).

\subsection{The D3~brane}\label{sec:D3}
\paragraph{RR four-form background II.}
Another interesting case to study is the \D{3}~brane. Since we would like to understand a four-dimensional worldvolume theory related to the \D5~brane in the previous section, we start from \eqref{eq:bg-D5}, and then apply T-duality twice in the $x^{2}$ and $x^{5}$ directions, respectively. The resulting background is almost the same, except that $x^{2}$ and $x^{5}$ appear interchanged:
\begin{equation}\label{eq:bg-D3}
  \begin{aligned}
    g_{mn}\dd{x^{m}}\dd{x^{n}} &= \Delta \left(\eta_{\alpha\beta} \dd{x^{\alpha}} \dd{x^{\beta}} + (\dd{x^{2}})^{2} \right)\\
    &\hspace{1.2in}  + \frac{\delta_{ab} \dd{x^{a}}\dd{x^{b}}+ (\dd{x^5})^2}{\Delta} + \left(\Delta \delta_{IJ}-\frac{U_{I}U_{J}}{\Delta}\right)\dd{x^{I}}\dd{x^{J}},\\
    C_{4} &= U \wedge \left(-\dd x^0\wedge \dd x^1\wedge \dd x^2  + \frac{\dd x^3\wedge \dd x^4\wedge \dd x^5}{\Delta^{2}}\right),
  \end{aligned}
\end{equation}
where  \(\alpha, \beta = 0,1\); \(a,b = 3,4\).
Again, $C_{4}$ can be also written down as
\begin{equation}
C_{4} =  -\frac{1}{4}\omega_{IJ}x^{J} \dd{x^{I}} \wedge \left( \Xi_{\alpha \beta i}\dd x^\alpha\wedge \dd  x^\beta  + \Xi_{ab i}\, \frac{\dd x^a\wedge \dd  x^b}{\Delta^{2}}\, \right) \wedge \epsilon^{i}{}_{j}\dd{x^{j}}\,,
\end{equation} 
where $\epsilon_{25} = 1$ for $i,j = 2,5$ and $\Xi$ takes the same values as in~\eqref{eq:xi-D5},
\begin{align}
\label{eq:xi-D3}
\Xi &= \frac{1}{2}\left(\Xi_{\alpha\beta 2 }\dd x^\alpha\wedge \dd x^\beta\wedge \dd x^2 + \Xi_{ab 5}\dd x^a\wedge \dd x^b\wedge \dd x^5\right)\nonumber\\
& = \frac{1}{2}\left(-\dd x^0\wedge \dd x^1\wedge \dd x^5 \pm \dd x^2\wedge \dd x^3\wedge \dd x^4\right).
\end{align}

\paragraph{Supersymmetric D3~brane action.}
T-duality in $x^{2}$ and $x^{5}$ simply dimensionally reduces the \D{5}~brane of the previous section to a \D{3}~brane extended in $\{x^{\mu}\}_{\mu=0,1,3,4}$. Thus, we can directly obtain the \D{3}~brane action via a \ac{kk} reduction on $\{x^{i}\}_{i=2,5}$ of the \D{5} action obtained in Section~\ref{sec:D5}.
The action contains again a twisted covariant derivative.
This time, however, the connection in the covariant derivative includes two transverse scalars and not the gauge field.
Writing the \ac{cs} term
\begin{equation}
S_{\text{cs}}^{\D3} =\frac{1}{g^{2}} \int \hat{C}_{4} =  -\frac{1}{g^{2}}\int \dd^{4}{x} \left( \frac{1}{\Delta^{2}}\Xi_{\alpha\beta5}\,\partial^{\beta}X^{5}\partial^{\alpha}X^{I} +\Xi_{ab2}\,\partial^{a}X^{2}\partial^{b}X^{I}  \right)\omega_{IJ}X^{J}\,,
\end{equation}
we see that the two scalars $X^{i}$ have to be included in the connections separately in the sectors $\{x^{\alpha}\}_{\alpha=0,1}$ and $\{x^{a}\}_{a=3,4}$:
\begin{equation}
\label{eq:connection-D3}
\mathcal{A}_{\mu} = \Xi_{\mu\nu i}\partial^{\nu}X^{i} =
\begin{cases}
  \frac{1}{2}\epsilon_{\alpha \beta} \del^\beta X^5 & \text{if \(\mu, \nu = \alpha, \beta = 0,1\)} \\
  \frac{1}{2}\epsilon_{ab} \del^b X^2 & \text{if \(\mu, \nu = a, b = 3,4\)}
\end{cases}
,
\end{equation}
which defines a twisted covariant derivative as before:
\begin{equation}
\label{eq:bcov-D3}
\mathcal{D}_{\mu}X^{I} = \partial_{\mu}X^{I}+\mathcal{A}_{\mu}\omega^{IJ}X_{J}\,.
\end{equation}
Thus, the bosonic action of the \D3~brane in $\{x^{\mu}\}_{\mu=0,1,3,4}$ takes the form
\begin{equation}
\label{eq:baction-D3}
\begin{aligned}
S_{B}^{\D3} =  -\frac{1}{g^{2}}\int \dd^{4}{x} \, \frac{1}{2}\sqrt{-g}g^{\mu\nu}g_{IJ}\mathcal{D}_{\mu}X^{I}\mathcal{D}_{\nu}X^{J} +  \frac{1}{2}\eta^{\mu\nu}\partial_{\mu}X^{i}\partial_{\nu}X^{j} + \frac{1}{4}\sqrt{-g}g^{\mu\rho}g^{\nu\sigma}F_{\mu\nu}F_{\rho\sigma}\,, 
\end{aligned}
\end{equation}
where $g_{\mu\nu}$ is the same metric as in \eqref{gbrane}.
Note that the coupling of the gauge field is different from what we had found for the six-dimensional system.
It does not appear anymore in the covariant derivative, but is a standard Maxwell term in curved space with metric \(g_{\mu\nu}\).

Let us move on to the fermionic action. As in the \D{5}~brane, the calculation shows that the deformation turns on the covariant derivative for the fermions as well as a Yukawa-like term in the Dirac action. Using a 32-component Majorana Weyl spinor $\Psi$, we find up to $\mathcal{O}(\varepsilon)$
\begin{equation}
  \label{eq:faction-D3}
  S_{F}^{\D3} = \frac{i}{2g^{2}}\int \dd[4]{x} \bar{\Psi} \bqty{ \Gamma^{\mu}\covD_{\mu} + \frac{1}{4}\Xi_{\mu\nu\rho}\Gamma^{\nu\rho}\omega_{IJ}\partial^{\mu}X^{I}\Gamma^{J}} \Psi + \mathcal{O}(\varepsilon^{2}) \,,
\end{equation}
where the covariant derivative on fermions is
\begin{equation}
\label{eq:fcov-D3}
\mathcal{D}_{\mu}\Psi = \partial_{\mu}\Psi + \mathcal{A}_{\mu}\frac{1}{4}\omega_{IJ}\Gamma^{IJ}\Psi\,.
\end{equation}
One cannot see any interaction between the fermion and $U(1)$ gauge field at the level of linear order $\mathcal{O}(\varepsilon)$ and suppressing higher derivatives.

\paragraph{Spacetime- and R-symmetries.}
 
Let us focus on the symmetries that the world volume action inherits from the ten-dimensional background. The presence of the \D3~brane in a flat background usually breaks the ten-dimensional Poincaré symmetry into two sectors: $SO(1,9) \rightarrow SO(1,3) \times SO(6)$, where the latter corresponds to the R-symmetry for the six transverse scalars. However, turning on the five-form flux triggers the covariant derivative~\eqref{eq:bcov-D3} as well as  the curved metric $g_{\mu\nu}$. As a result, the $SO(1,3)$ worldvolume symmetry in the action~\eqref{eq:baction-D3} explicitly splits into $SO(1,1) \times SO(2)$. This is expected to be ascribed to the presence of surface defects, living on $\{x^{\alpha}\}_{\alpha=0,1}$ and $\{x^{a}\}_{a=3,4}$, respectively.

It makes sense to carry out the analysis by dimensionally reducing the 4d theory on a torus generated by $\{x^{a}\}_{a=3,4}$. The resulting theory will be $d=2, \mathcal{N}=(4,4)$, as there are 8 Killing spinors preserved on the \D3 world volume as shown later. 

As for the global symmetry for scalars, recall that two scalars $X^{i}$ enter the connection $\mathcal{A}_{\mu}$ in~\eqref{eq:connection-D3} and they decouple from the sextuplet for the original $SO(6)$ R-symmetry group.
Thus, the R-symmetry acts only on the four transverse scalars $X^{I}$.
Using the same argument as in the \D{5}~brane case, we see that the R-symmetry $SO(6)$ is broken to $SU(2)$ by turning on the five-form flux.
A manifestly \(SU(2)\)-invariant action can be written by introducing a complex doublet $\mathbf{Z}$ given in~\eqref{eq:doublet-D5}.

Finally, it may be interesting to see the effect of S-duality.
Recall that no dilaton or Kalb--Ramond  field is turned on in the flux background \eqref{eq:bg-D3}.
Therefore, both the \D{3}~brane and the background configuration map to themselves, which means that the gauge theory in presence of the defects remains invariant like in the undeformed \(\mathcal{N} = 4\) theory.

\paragraph{Equations of Motion.}
The D3 brane action possesses a reduced Lorentz symmetry $SO(1,1)$ on the $\{x^{\alpha}\}$-plane. This motivates us to use light-cone coordinates $(x^{+},x^{-})$ for the $\{x^{\alpha}\}$-plane 
\begin{equation}
x^{\pm} = x^{0} \pm x^{1}
\end{equation}
and complex coordinates $(\sigma, \bar{\sigma})$ for the $\{x^{a}\}$-plane
\begin{equation}
\sigma = x^{3} + i x^{4}\,,\qquad \bar{\sigma} = x^{3} - i x^{4}\,.
\end{equation}
Using these coordinates, we vary the action \eqref{eq:baction-D3} with respect to $\mathbf{Z}^{\dagger}$ \eqref{eq:doublet-D5}, and find
\begin{equation}
\label{eq:4dEOMZ}
\begin{aligned}
&\{\mathcal{D}_{+},\mathcal{D}_{-}\}\mathbf{Z} + i\varepsilon \left(\left\{\mathcal{D}_{+},\frac{\mathcal{J}_{-}}{\Delta^{2}}\right\} + \left\{\mathcal{D}_{-}, \frac{\mathcal{J}_{+}}{\Delta^{2}}\right\}\right)\mathbf{Z} - \frac{2\varepsilon^{2}}{\Delta^{4}}\mathcal{J}_{+}\mathcal{J}_{-}\mathbf{Z}\\
&\qquad -\left(\mathcal{D}_{\sigma}(\Delta^{2}\mathcal{D}_{\bar{\sigma}}\mathbf{Z}) +\mathcal{D}_{\bar{\sigma}}(\Delta^{2}\mathcal{D}_{\sigma}\mathbf{Z}) \right) - i\varepsilon \left(\{\mathcal{D}_{\sigma}, \mathcal{J}_{\bar{\sigma}}\} +\{\mathcal{D}_{\bar{\sigma}}, \mathcal{J}_{\sigma}\}\right)\mathbf{Z}\\
&\qquad\qquad+ \varepsilon^{2} \left( (\mathcal{D}_{\sigma}\mathbf{Z})^{\dagger}(\mathcal{D}_{\sigma}\mathbf{Z}) + (\mathcal{D}_{\bar{\sigma}}\mathbf{Z})^{\dagger}(\mathcal{D}_{\bar{\sigma}}\mathbf{Z})\right)\mathbf{Z}\\
&\qquad + \frac{\varepsilon^{2}}{2}\left(F_{+-}F^{+-} - \frac{1}{\Delta^{4}}F_{\sigma\bar{\sigma}}F^{\sigma\bar{\sigma}}\right)\mathbf{Z} = 0\,,
\end{aligned}
\end{equation}
where $\{\cdot,\cdot\}$ is a conventional anti-commutator and $\mathcal{J}$ is as in~\eqref{eq:Current}.
For the other real scalars $X^{2}$ and $X^{5}$, we obtain
\begin{equation}
(\partial_{+}\partial_{-} - \partial_{\sigma}\partial_{\bar{\sigma}})X^{2} + i \left(\partial_{\sigma}\mathcal{J}_{\bar{\sigma}} - \partial_{\sigma}\mathcal{J}_{\bar{\sigma}}\right) = 0
\end{equation}
and 
\begin{equation}
(\partial_{+}\partial_{-} - \partial_{\sigma}\partial_{\bar{\sigma}})X^{5} + i \left(\partial_{+}\left(\frac{\mathcal{J}_{-}}{\Delta^{2}}\right) - \partial_{-}\left(\frac{\mathcal{J}_{+}}{\Delta^{2}}\right)\right) = 0 .
\end{equation}
For the gauge fields, the \ac{eom} are 
\begin{equation}
\begin{aligned}
\partial_{k} (\Delta^{2} F^{kl}) + \partial_{p}F^{pl} = 0\,,\\
\partial_{p} \left(\frac{1}{\Delta^{2}} F^{pq}\right) + \partial_{k}F^{kq} = 0\,,
\end{aligned}
\end{equation}
where $k,l=+,-$ and $p,q = \sigma, \bar{\sigma}$. The derivatives $\partial_{\pm}$ are associated to $x^{\pm}$, respectively.

Let us consider a 1+1 dimensional sector inside the four-dimensional worldvolume.  Suppose that all the fields are dependent only on $x^{\pm}$. Then, for example, \eqref{eq:4dEOMZ} is reduced to a very compact form:
\begin{equation}
\left\{\mathcal{D}_{+} + i\varepsilon\frac{\mathcal{J}_{-}}{\Delta^{2}}, \mathcal{D}_{-} + i\varepsilon \frac{\mathcal{J}_{+}}{\Delta^{2}}\right\}\mathbf{Z}  + \frac{\varepsilon^{2}}{2}F_{+-}F^{+-}=0\,.
\end{equation}
In addition, we can express the other equations of motion compactly via differential forms. The equation for $X^{2}$ is the equation for a free field, as $X^{2}$ does not enter the covariant derivative due to the restriction to two dimensions:
\begin{equation}
\dd \star_{2} \dd X^{2} = 0\,,
\end{equation}
which can be seen as the equation for a conserved current $\dd X^{2}$. 
On the other hand, the equation for $X^{5}$ takes the form
\begin{equation}
\label{eq:2dEOMX5}
\dd \star_{2} \left(\dd X^{5} - \star_{2} \frac{2}{\Delta^{2}}\mathcal{J} \right) = \dd \left( \star_{2} \dd X^{5} - \frac{2}{\Delta^{2}}\mathcal{J}\right) = 0\,.
\end{equation} 
We thus find locally a free scalar $\phi$ satisfying 
\begin{equation}
\dd \phi = \star_{2}\dd X^{5} - \frac{2}{\Delta^{2}}\mathcal{J}\,.
\end{equation}
Using the connection $\mathcal{A}_{\alpha}$ in \eqref{eq:connection-D3}, one finds 
\begin{equation}
\partial_{\alpha} \phi = -2 \mathcal{A}_{\alpha} + \frac{2}{\Delta^{2}}\mathcal{J}_{\alpha}\,,
\end{equation}
In analogy to our result in~\eqref{eq:dualscalarD5}.
Alternatively, \eqref{eq:2dEOMX5} can be interpreted as a conservation law for the current $\mathcal{K}$:
\begin{equation}
\mathcal{K} = \dd X^{5} - \star_{2}\frac{2}{\Delta^{2}}\mathcal{J} = -2 \star_{2} (\mathcal{A} + \frac{1}{\Delta^{2}}\mathcal{J})\,.
\end{equation}
Finally, the equation for the gauge fields labeled by $\sigma, \bar{\sigma}$ becomes free:
\begin{equation}
\dd\star_{2} \dd A^{p} = 0\,,\quad p=\sigma, \bar{\sigma}
\end{equation}
which implies that $\partial_{-}\partial_{+}A^{p} = 0$. The other gauge field $A^{\pm}$ is subject to the deformation: 
\begin{equation}
\dd \star_{2} \left(\Delta^{2}F \right) = 0\,,
\end{equation}
where the field strength is restricted with the only non-vanishing components being $F_{+-} = \partial_{+}A_{-} - \partial_{-}A_{+}$. Due to the dimensionality, we obtain
\begin{equation}
\Delta^{2} \star_{2} F = c_{0} \equiv const.
\end{equation}
Consequently, for the gauge field $A^{\pm}$, we have
\begin{equation}
\star_{2}\dd A = c_{0} \Delta^{-2}\,.
\end{equation}
\paragraph{Supersymmetry.}
The discussion of the supersymmetry goes along the same lines as for the \D5 brane. It follows from the gravitino equation that 16 Killing spinors, preserved on a classical \D3~brane, are reduced by half on the \D3~brane in the flux background~\eqref{eq:bg-D3}. The supersymmetry transformations take almost the same form except for two scalars $X^{j}, j=2,5$. We obtain up to the first order $\mathcal{O}(\varepsilon)$
\begin{equation}
\label{eq:susy-D3}
  \begin{aligned}
    \delta X^{j} &= i \bar{\epsilon}(X) \Gamma^{j}\Psi\\
    \delta X^{I} &= i \bar{\epsilon}(X) \Gamma^{I}\Psi\\    
    \delta A_{\mu} &= i \bar{\epsilon}(X) \Gamma_{\mu}\Psi\\
    \delta \Psi  &= \pqty{\Gamma^{\mu j}\partial_{\mu}X^{j} + \Gamma^{\mu I}\partial_{\mu}X^{I} + \frac{1}{2}F_{\mu\nu}\Gamma^{\mu\nu}} \epsilon(X)\,,
\end{aligned}
\end{equation}
where the non-constant 32-component Killing spinor $\epsilon(x)$ is expressed as
\begin{equation}
\label{eq:Killing-D3}
\epsilon(X) = \left(1 +  \frac{1}{8}\Xi_{\mu\nu i}\Gamma^{\mu\nu i}\omega_{IJ}X^{I}\Gamma^{J}\right)\epsilon\,.
\end{equation}
In solving the gravitino equation, we obtain a first-order constraint on the constant spinor $\epsilon$
\begin{equation}
\omega_{IJ}\Gamma^{IJ}\epsilon=0\,,
\end{equation}
which implies 8 independent spinors as expected. 

\paragraph{Non-Abelian Generalisation.}
Finally, we make a short remark on a stack of \D3~branes based on the non-Abelian \D5~brane action ~\eqref{Sna}. Applying the dimensional reduction to both $x^{2}$ and $x^{5}$ directions, we can naturally derive the non-Abelian \D3~brane action:
\begin{equation}
\label{eq:nonAbeaction-D3}
\begin{aligned}
S &= -\frac{1}{g^{2}}{\rm tr}\int \dd^{4}{x} \frac{1}{4}F_{\mu\nu}F^{\mu\nu} + \frac{1}{2}D_{\mu}X^{j}D^{\mu}X^{j} + \frac{1}{2}D_{\mu}X^{I}D^{\mu}X^{I} \\
&\hspace{2cm}-\frac{1}{2}[X^{j},X^{I}][X^{j},X^{I}] - \frac{1}{4}[X^{j},X^{k}][X^{j},X^{k}]- \frac{1}{4}[X^{I},X^{J}][X^{I},X^{J}]\\
&-\frac{1}{g^{2}}{\rm Str} \int \dd^{4}{x} \frac{i}{2}\bar{\Psi}\Gamma^{\mu}D_{\mu}\Psi + \frac{1}{2}\bar{\Psi}\Gamma^{j}[X^{j},\Psi] + \frac{1}{2}\bar{\Psi}\Gamma^{I}[X^{I},\Psi]\\
&\quad + \Xi_{\mu\nu j}D^{\nu}X^{j}\omega^{IJ}X^{J} [ X^{j},X^{I}] - \frac{i}{16}\bar{\Psi}\Gamma^{j}\Xi_{j\mu\nu}F^{\mu\nu}\omega_{IJ}\Gamma^{IJ}\Psi-\frac{i}{8}\bar{\Psi}\Gamma^{\mu}\Xi_{\mu\nu j}D^{\nu}X^{j}\omega_{IJ}\Gamma^{IJ}\Psi\\
&-\frac{i}{8}\bar{\Psi}D^{j}X^{I}\Gamma^{J}\omega_{IJ}\Xi_{j\mu\nu}\Gamma^{\mu\nu}\Psi - \frac{i}{4}\bar{\Psi}D^{\mu}X^{I}\Gamma^{J}\omega_{IJ}\Xi_{\mu\nu j}\Gamma^{\nu j}\Psi + \frac{3}{2}\bar{\Psi}\Xi_{\mu\nu j}\Gamma^{\mu\nu j}\omega_{IJ}[X^{I},X^{J}]\Psi\,,
\end{aligned}
\end{equation}
where the reduced directions are labeled by $i,j=2,5$ again. 
The corresponding Killing spinor is obtained by reducing (\ref{eq:nonAbe-Killingspinor}) in the same way: 
\begin{equation}
\epsilon^{a}{}_{b}(X) = \delta^{a}_{b}\epsilon + \frac{1}{8}\Xi_{\mu\nu j }\Gamma^{\mu\nu j} X^{J}_{c}\omega^{JK}\Gamma^{K}d^{ac}{}_{b}\epsilon\,.
\end{equation}

\subsection{The duality cascade} \label{sec:Duality-Cascade}

 \newcommand*{\dirA}{\tikz[baseline=-0.5ex]\draw[black,fill=black,radius=2pt] (0,0) circle ;}%
\newcommand*{\dirB}{\tikz[baseline=-0.5ex]\draw[black,radius=2pt] (0,0) circle ;}%

\begin{figure}
 \centering
   \begin{tikzcd}
                     & \D4                   & \D3                             & \D2 & \D1                               \\
                     &                       & {\dirA \ \dirA } \ar[r]\ar[ddr]       & {\dirA \ \dirA \ \dirA} \ar[dr]         \\
                     & \dirA \ar[ur] \ar[dr] &                                 &     & {\dirA \ \dirA \ \dirA \ \dirB}   \\
 \D5 \ar[ur] \ar[dr] &                       & {\dirA \ \dirB} \ar[r] \ar[ddr] & {\dirA \ \dirA \ \dirB} \ar[ur] \ar[dr] \\
                     & \dirB \ar[ur] \ar[dr] &                                 &     & {\dirA \ \dirA \ \dirB \ \dirB}   \\
                     &                       & {\dirB \ \dirB} \ar[r]          & {\dirB \ \dirB \ \dirA} \ar[ur]         \\
   \end{tikzcd}
 \caption{Duality web of the theory on the \D5 brane. A filled dot  indicates a T-duality in one of the directions \(x^2, x^3, x^4\), an empty dot  indicates a T-duality in the direction \(x^0, x^1, x^5\) (although we don't consider T-duality along $x^0$) The order of the dots is irrelevant because T-duality is commutative.  }
 \label{fig:duality-web-D5}
\end{figure}
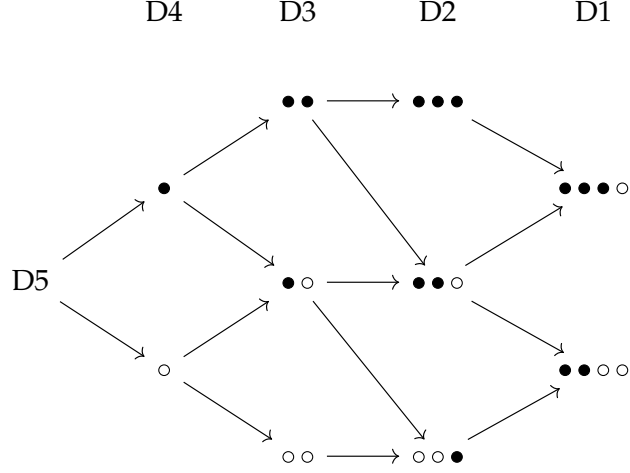

\newcommand*{\splitrow}[2]{\begin{tabular}{@{}r@{}}#1\\[.1cm]#2\end{tabular}}
\begin{table}[t]
 \centering
 \begin{tabular}{llrrc}
   \toprule
   && directions& \({\cal A}\) & plane \\
   \midrule
   \D5 & & 0,1,2,3,4,5 & \splitrow{\(\star_3F     \)}{\(\star_3F \)}& \splitrow{\(\dd{x^0} \wedge \dd{x^1} \wedge \dd{x^5} \)}{\(\dd{x^2} \wedge \dd{x^3} \wedge \dd{x^4}\)} \\[.5cm]
   \midrule
   \multirow{4}{*}{\D4} & \dirA & 0,1,3,4,5 & \splitrow{\(\star_3F \)}{\(\star_2\dd{X^2} \)} & \splitrow{\(\dd{x^0} \wedge \dd{x^1}\wedge \dd{x^5}\)}{\(\dd{x^3} \wedge \dd{x^4} \)}  \\[.5cm]
   & \dirB & 0,1,2,3,4 & \splitrow{\(\star_2\dd{X^5} \)}{\(\star_3F \)} & \splitrow{\(\dd{x^0} \wedge \dd{x^1} \)}{\(\dd{x^2} \wedge \dd{x^3}\wedge \dd{x^4}\)}  \\[.5cm]
   \midrule
    & \dirA \ \dirA & 0,1,3,5  & \splitrow{\(\star_3F \)}{--} & \splitrow{\(\dd{x^0}\wedge \dd{x^1}\wedge \dd{x^5}\)}{\(\dd x^3\)}  \\[.5cm]
   \D3 & \dirA \ \dirB  & 0,1,3,4 & \splitrow{\(\star_2\dd{X^5} \)}{\(\star_2\dd{X^2}  \)} &\splitrow{\(\dd x^0\wedge \dd x^1 \)}{\(\dd x^3\wedge \dd x^4\)}   \\[.5cm]
   & \dirB \ \dirB & 0,2,3,4 & \splitrow{--}{\(\star_3F \)} & \splitrow{\(\dd x^0 \)}{\(\dd x^2\wedge \dd x^3\wedge \dd x^4\)}   \\[.5cm]
   \midrule
    & \dirA \ \dirA \ \dirA  & 0,1,5 & \(\star_3F \) & \(\dd x^0\wedge \dd x^1\wedge \dd x^5\)  \\[.5cm]
   \D2 & \dirA \ \dirA \ \dirB & 0,1,3 & \splitrow{\(\star_2\dd{X^5} \)}{--}& \splitrow{\(\dd x^0\wedge \dd x^1\)}{\(\dd x^3\)} \\[.5cm]
   & \dirB \ \dirB \ \dirA & 0,3,4 &\splitrow{--}{\(\star_2\dd{X^5} \) }& \splitrow{\(\dd{x^0}\)}{\(\dd x^3\wedge \dd x^4\)}  \\[.5cm]
   \midrule
   \multirow{2}{*}{\D1} & \dirA \ \dirA \ \dirA \ \dirB & 0,1 & \(\star_2\dd{X^2} \) & \(\dd{x^0} \wedge \dd{x^1}\)  \\[.5cm]
    & \dirA \ \dirA \ \dirB \ \dirB & 0,3 & -- & \(\dd{x^0} \wedge \dd{x^3}\) \\
   \bottomrule
 \end{tabular}
 \caption{Duality web of brane deformations starting from the \D{5}~brane setup. We give the type of T-duality as in Fig.~\ref{fig:duality-web-D5} in the second column, the form of the deformation connection ${\cal A}$   in the third column  along with the plane in which that ${\cal A}$ acts in the fourth column (note that we have dropped any signs for clarity).}
 \label{tab:duality-web-D5}
\end{table}

So far we have studied \D{5}~branes and \D{3}~branes in flux backgrounds leading to novel deformations in terms of a twisted covariant derivative.  In particular the deformations take two forms: one involving the gauge field strength and one the derivative of the scalars
\begin{equation}\label{As}
  \begin{aligned}
    \D5: &&  {\cal A}_\mu =\frac 12 \Xi_{\mu\nu\lambda}F^{\nu\lambda} , \\
    \D3: && {\cal A}_\mu = \Xi_{\mu\nu i}D^{\nu}X^i .
  \end{aligned}
\end{equation}
More conceptually we can think of these as follows. In the  \D5~brane deformation splits the worldvolume into two planes: \(x^2, x^3, x^4\) and \(x^0, x^1, x^5\). In each of these planes ${\cal A} \sim \star_3 F$ where $\star_3$ is the associated 3-dimensional Hodge dual. Upon reduction to the D3~brane we find two 2-dimensional planes and ${\cal A}\sim \star_2 dX$ where   $\star_2 $ refers to the appropriate 2-dimensional Hodge dual and $X$ is either $X^2$ or $X^5$. So roughly speaking we can think of as (\ref{As}) as
\begin{equation}
  \begin{aligned}
    \D5: &&  {\cal A} \sim \star_3 F , \\
    \D3: && {\cal A}  \sim \star_2 \dd X^{i}\, .
  \end{aligned}
\end{equation}

The \D{3}~brane case is obtained from the \D{5}~branes by T-duality which on the worldvolume is simply dimensional reduction. 
However there are many other examples that are related by T-duality. In order to preserve supersymmetry we require $\omega_{67}=\pm\omega_{89}$ to be non-vanishing and hence we can only perform T-dualities along $x^1,...,x^5$ (we do not consider a time-like T-duality). Rather than detail each case we simply wish to list the possibilities. The exact form of the action can be obtained by dimensional reductions of the \D5~brane action we constructed above. In each of the cases typically both types of covariant derivative appear but in different subplanes of the worldvolume.

In Figure \ref{fig:duality-web-D5} we list the duality cascade that 
arises depending on which directions are T-dualized.  In particular the original \D5~brane deformation splits the worldvolume into two planes: \(x^2, x^3, x^4\) and \(x^0, x^1, x^5\). A filled dot denotes a T-duality in the first plane and an empty dot a T-duality in the second plane.   The structure of the worldvolume theory deformation of the inequivalent configurations we can reach starting from the \D5~brane are collected in Table~\ref{tab:duality-web-D5}.

\section{Conclusions}
\label{sec:conclusions}

In this paper we have constructed and studied various supersymmetric deformations of non-Abelian gauge theories derived from String Theory by putting \D{p}~branes into flux backgrounds. In the first case the deformation takes the form of a Wilson line for a connection that twists the R-symmetry with the gauge algebra. It can be easily constructed for any gauge theory as an exact deformation. We explicitly presented it for Yang--Mills gauge theories as well as the Chern--Simons-matter theories on \M2~branes, including a maximally supersymmetric case that preserves all the supersymmetries of the \ac{abjm} model.  

We also constructed a higher-dimensional and higher order-analogue, first for \D5~branes but then reduced it to \D3~branes and other \D{p}~branes. This deformation also twists the R-symmetry into the gauge symmetry but with a non-trivial connection. It induces higher-derivative corrections to the gauge theory while preserving half of the supersymmetry. In these cases we have only been able to construct the non-Abelian theory and supersymmetry to first order in the deformation. It would be interesting to extend our analysis to the next order. In particular it would be important to see whether or not the twisted covariant derivative structure persists. 

The first examples have a clear interpretation as the insertion of a Wilson-line defect into the gauge theory. 
For the higher-order deformations  the Lorentz symmetry of the underlying gauge theory is broken from \(SO(1, p)\) to \(SO(1,(p-1)/2) \times SO((p-1)/2)\) and one would be tempted to associate the deformation to an extended \((p-1)/2\)-dimensional defect or possibly two intersecting defects. Other \(2\)-dimensional gauge theory defects have appeared in~\cite{Gukov:2006jk, Gukov:2008sn, Gaiotto:2014ina}. However, our deformations preserve different global symmetries and, like the Omega-deformation, are higher order in the fields. As such they cannot  easily  be identified with these other examples discussed in the literature. 

It would also be interesting to see if one could relate these higher-order deformations to defects associated to the two-group symmetries that have appeared recently in~\cite{Gaiotto:2014kfa,Cordova:2018cvg}. It is also intriguing to note that a similar kind of twisted covariant derivative, where the connection is given in terms of a field strength, has also appeared recently in the work~\cite{Ganor:2017rrz} in relation to non-local descriptions of the \M5~brane. Again it would be interesting to see if there is a deeper relation, with our deformation arising in the local limit.

\section*{Acknowledgements}
The authors would would like to thank Chris Hull and Kimyeong Lee for useful discussions in an early stage of this project and more recently Ori Ganor and Shigeki Sugimoto. 

Y.S. is also grateful to the Yukawa Institute for Theoretical Physics at Kyoto University, where part of this work was developed during the YITP-W-17-07 on Strings and Fields 2017. D.O. and S.R. would like to acknowledge support from the Simons Center for Geometry and Physics, Stony Brook University at which some of the research for this paper was performed.

N.L. was supported in part by \textsc{stfc} grant \textsc{st/p000258/1}. 
D.O. %
acknowledges partial support  by the \textsc{nccr 51nf40-141869} ``The Mathematics of Physics'' (SwissMAP).
The work of S.R. and Y.S. is supported by the Swiss National Science Foundation (\textsc{snf}) under grant number \textsc{pp00p2\_157571/1}.

\appendix

\section{Duality web of the string backgrounds}
\label{sec:Melvin}

In this appendix we explain how the \ac{rr} four-form backgrounds~\eqref{eq:bg-D5} and~\eqref{eq:bg-D3} can be constructed by starting from a very simple set-up: a flat background with Melvin identifications, where no dilaton or Kalb--Ramond two-form $B$ is
turned on. The various steps are   listed in Table \ref{tab:probe-branes}.

The starting point is a locally-flat space with ``Melvin identifications''.
This is a non-trivial fibration over a circle \(u\) (the Melvin direction) of the type
\begin{equation}
  \begin{cases}
    u \simeq u + 2 \pi R_u,\\
    \theta_A \simeq \theta_A + \varepsilon_A R_u & \text{for \(A = 1,2,3,4\)},
  \end{cases}
\end{equation}
where \(R_u\) is the radius of the direction \(u\), \(\theta_A\) is the angle in the plane spanned by \(x^{2A}\) and \(x^{2A + 1}\) and the \(\varepsilon_A\) are real parameters.
Decoupling the circles and T-dualizing in \(u\) we obtain the \emph{fluxtrap} background~\cite{Hellerman:2011mv,Orlando:2013yea} where the non-trivial fibration is traded for a curved spacetime, a \(B\)-field and a dilaton.
For simplicity we consider only two non-vanishing \(\varepsilon\) parameters and we identify \(u\) with \(x^3\) to find:
\begin{equation}
\label{eq:bg-IIA1}
  \begin{aligned}
    g_{mn}\dd{x^{m}}\dd{x^{n}} &= -(\dd x^0)^2 +(\dd x^1)^2+(\dd x^4)^2+(\dd x^5)^2   + \frac{(\dd{x^{3}})^{2}}{\Delta^{2}} \\
    & \qquad + \left(\delta_{IJ} - \frac{U_{I}U_{J}}{\Delta^{2}}\right)dx^{I}dx^{J},\\
       B_{2} &= U \wedge \frac{dx^{3}}{\Delta^{2}}\,,\\
       e^{\Phi} &= \Delta^{-1}\,,
     \end{aligned}
\end{equation}
where $I, J = 6,7,8,9$, $U = \frac{1}{2}\omega_{IJ}x^{I}\dd{x^{J}}$, and $\Delta=\sqrt{1+U_{I}U^{I}}$. 

The backgrounds that we use in this work are ``S-dual'' to this one (that we think of as a \tIIA configuration).
More precisely we need to perform a 9-11 flip: we lift~\eqref{eq:bg-IIA1} by adding $x^{10}$ and then reduce on $x^{4}$.
The result is a \tIIA background with a \ac{rr} three-form potential~\cite{Hellerman:2012zf}:
\begin{equation}
\label{eq:bg-IIA2}
  \begin{aligned}
    g_{mn}\dd{x^{m}}\dd{x^{n}} &= \Delta \left(-(\dd x^0)^2 +(\dd x^1)^2+(\dd x^2)^2+(\dd x^5)^2   \right) + \frac{  (\dd x^3)^2+(\dd x^{10})^2  }{\Delta} \\
    &\qquad  + \left(\Delta \delta_{IJ} - \frac{U_{I}U_{J}}{\Delta}\right)dx^{I}dx^{J},\\
     C_{3} &=U \wedge \frac{\dd{x^{3}}\wedge \dd{x^{10}}}{\Delta^{2}},\\
    e^{\Phi} &= \Delta^{1/2}.
  \end{aligned}
\end{equation}

There are two inequivalent ways to dualise to \tIIB.
First, we can apply a T-duality in $x^{2}$ to obtain the background~\eqref{eq:bg-D5} in which the \D5~brane of Section~\ref{sec:D5} lives:
\begin{equation}
  \begin{aligned}
  \label{eq:bg-D52}
  g_{mn}\dd{x^{m}}\dd{x^{n}} &= \Delta \left( -(\dd x^0)^2 +(\dd x^1)^2+(\dd x^5)^2  \right) + \frac{ (\dd x^2)^2 +(\dd x^3)^2+(\dd x^{10})^2}{\Delta} \\
  &\qquad + \left(\Delta \delta_{IJ}    - \frac{U_I U_J}{\Delta}\right) \dd{x^{I}} \dd{x^{J}},\\
    C_{4} &= U \wedge \left( -\dd{x^{0}}\wedge \dd{x^{1}}\wedge \dd{x^{5}} + \frac{\dd{x^{2}}\wedge\dd{x^{3}}\wedge \dd{x^{10}}}{\Delta^{2}}\right).
  \end{aligned}
\end{equation}
Alternatively, a T-duality in $x^{5}$ in~\eqref{eq:bg-IIA2} leads us to the other \ac{rr} four-form background \eqref{eq:bg-D3} used for the \D3~brane in Section~\ref{sec:D3}:
\begin{equation}
  \begin{aligned}
  \label{eq:bg-D32}
    g_{mn}\dd{x^{m}}\dd{x^{n}} &= \Delta \left(-  (\dd x^0)^2 +(\dd x^1)^2+(\dd x^5)^2  \right) + \frac{   (\dd x^3)^2 +(\dd x^{10})^2+(\dd x^5)^2  }{\Delta} \\
    &\qquad +  \left(\Delta \delta_{IJ}    - \frac{U_I U_J}{\Delta}\right) \dd{x^{I}} \dd{x^{J}},\\
    C_{4} &= U \wedge \left( \dd{x^{0}}\wedge \dd{x^{1}}\wedge \dd{x^{2}} + \frac{\dd{x^{3}}\wedge\dd{x^{10}}\wedge \dd{x^{5}}}{\Delta^{2}}\right)\,.
  \end{aligned}
\end{equation}
Note that the oxidised coordinate $x^{10}$ is renamed to $x^{4}$ in Section~\ref{sec:higher-dimensional-defects}.

\begin{table}
\centering
\begin{tabular}{ccccccccccccc}
\toprule
background                          & probe branes & $0$ & $1$ & $2$ & $3$ & $4$            & $5$ & $6$                               & $7$                                  & $8$ & $9$ & $10$           \\
                                    &              &     &     &     &     &                &     & \multicolumn{2}{c}{$\varepsilon$} & \multicolumn{2}{c}{$\mp\varepsilon$} &                            \\[-0.08cm]
\midrule
locally flat~\textsc{iib}\xspace    & \D3          & \X  & \X  & \X  &     & \X             &     &                                   &                                      &     &     & $\blacksquare$ \\[.1cm]
\eqref{eq:bg-IIA1}                  & \D4          & \X  & \X  & \X  & \X  & \X             &     &                                   &                                      &     &     & $\blacksquare$ \\[.1cm]
\eqref{eq:bg-IIA2}                  & \D4          & \X  & \X  & \X  & \X  & $\blacksquare$ &     &                                   &                                      &     &     & \X             \\[.1cm]
\eqref{eq:bg-D52}, \eqref{eq:bg-D5} & \D5          & \X  & \X  & \X  & \X  & $\blacksquare$ & \X  &                                   &                                      &     &     & \X             \\[.1cm]
\eqref{eq:bg-D32}, \eqref{eq:bg-D3} & \D3          & \X  & \X  &     & \X  & $\blacksquare$ &     &                                   &                                      &     &     & \X             \\[.1cm]
\bottomrule 
\end{tabular}
\caption{Probe branes and their backgrounds. A black square $\blacksquare$ means that the direction is not part of the ten-dimensional geometric description.}
\label{tab:probe-branes}
\end{table}

\section{Notation}
\label{sec:notation}

We  use the convention that spacetime coordinates have lower-case symbols \emph{e.g.} $x^I$, $z^A$ whereas the corresponding scalar fields on the \D~brane  have upper-case symbols \emph{e.g.} $X^I$, $Z^A$.
 
Throughout the paper  we have specified the range of the indices in each of the different sections, but the general rule for the notation of the coordinate indices is:
\medskip

\begin{tabular}[c]{ll}
  \(m,n,p\) & the ten-dimensional bulk\\
  \(I,J,K\) & transverse directions to a  brane with \(\omega_{Im}\ne 0\)\\
  \(\mu ,\nu,\rho\) & worldvolume coordinates of a   brane\\
  \(\alpha,\beta,\gamma\) & a subspacetime on  a  brane\\
  \(a,b,c\) & the   orthogonal subspace on  a  brane.\\  
\end{tabular}

\medskip

For the \D3~brane we introduce another set of indices:
\medskip

\begin{tabular}[c]{ll}
\(i,j,k = 2,5\) & transverse directions to the brane  with \(\omega_{im}= 0\).
\end{tabular}

\setstretch{1}

\printbibliography

\end{document}